\documentclass[useAMS,usenatbib]{mn2e}
\usepackage{epsfig}
\usepackage{graphicx}
\usepackage{amssymb}
\usepackage{color}
\usepackage{placeins}
\usepackage{subfigure}
\usepackage[fleqn]{amsmath}
\begin{document}
%%%%%%%%%%%%%%%%%%%%%%%%%%%%%%%%%%%%%%%%%%%%%%%%%%%%%%%%%
\title[Studies of EBL with TeV BL Lacs]
  {Studies of Extragalactic Background Light with TeV BL Lac Objects}

\author[Qin et al.]
  {Qin, Longhua$^{1}$,
    Wang, Jiancheng$^{2,3}$, Gao, Quangui$^{1}\thanks{qggao@yxnu.edu.cn}$, Na, Weiwei$^{1}$,  \\
     \newauthor Li, Huaizhen$^{1}$, Ma, Ju$^{1}$, and Yang, Jianping$^{4}$\\
     $^1$Physics Department, Yuxi Normal University, Yuxi, Yunnan, 653100, China\\
   $^2$Yunnan Observatory, Chinese Academy of Sciences, Kunming,
Yunnan, 650011, China. \\
    $^3$Key Laboratory for the Structure and Evolution of Celestial Objects, Chinese Academy of Sciences,  Kunming, Yunnan, 650011, China.\\
   $^4$Yunnan Agricultural University, Kunming,Yunnan, 650201, China}

%\pagerange{\pageref{firstpage}--\pageref{lastpage}} \pubyear{2018}

\maketitle

\label{firstpage}

%%%%%%%%%%%%%%%%%%%%%%%%%%%%%%%%%%%%%%%%%%%%%%%%%%%%%%%%%
\begin{abstract}

Very high energy (VHE; $E \geq$ 100GeV) $\gamma-$ray from cosmological distances are attenuated by the extragalactic background light (EBL) in the IR to UV band. By contrasting measured versus intrinsic emission, the EBL photon density can be derived. However, neither the intrinsic spectra nor the EBL is separately known~-~except their combined effect. Here we first present a flexible model-dependent optical depth method to study EBL by fitting the emission spectra of TeV BL Lacs objects via a one-zone leptonic synchrotron self-Compton model (SSC). We have little information about electron energy distributions (EEDs) in the jet, which is critically important to build spectral energy distributions (SEDs) in SSC scenario. Based on the current particle acceleration models, we use two types of EEDs, i,e., a power-law log-parabola (PLLP) EED and a broken power-law (BPL) EED to fit the observed spectra. We obtain that the upper limits of EBL density are around about 30 n W m$^{-2}$ sr$^{-1}$, which are similar to the published measurement. Furthermore, we propose an unprecedented method to test radiation mechanisms involved in TeV objects by simply comparing the reduced EBL density with the limit obtained by galaxy counts, and demonstrate that for some BL Lacs, at least, the one-zone SSC model should be reconsidered.

\end{abstract}


%--------------------------------------------------------------------------------------------------------------------------------------------------------

\section{INTRODUCTION}
The diffuse extragalactic background light (EBL) from far-infrared to ultraviolet, is thought to be the radiation accumulated in the history of the Universe, dominated by starlight, either through direct emission or absorption/re-emission by dust. The spectral energy distributions (SEDs) of EBL marked as the two humps, their first one is arising from starlight and peaking at $\lambda\sim 1\mu m$, and the second one is locating in the warm dust and peaking at $\lambda\sim 100\mu m$. The EBL contains the critical information of the star formation processes and galaxy evolution. 

The direct measurement still is hardly achieved due to the zodiacal foreground light \citep{hau1998}, so the level of obtained EBL density is uncertain by a factor of several. The another approach to estimate the EBL is arising from the galaxy counts by deep optical/infrared observations \citep{mad2000,faz2004,dri2016}. This method may miss the faint undetected sources as well as the outer regions of the normal galaxies, hence it is thought to be the lower limits of EBL  density \cite[e.g.][]{ber2002}.

BL Lac objects with very high energy gamma-ray emission (VHE; $\geq$100 GeV) have been used to constrain the EBL. Their observed emission is jet dominated and the observed SEDs are bimodal. Generally, the first bump from radio to UV or X-rays is explained with the synchrotron emission of relativistic electrons accelerated in the jets, and the second bump at the GeV-TeV band is probably originated from the inverse Compton (IC) scattering of the same electron population. Their VHE gamma-rays are attenuated by low energy EBL photons through the pair production process ($\gamma_{\rm {VHE}}\gamma_{\rm {EBL}}\rightarrow e^+e^-$) . With assumptions about intrinsic gamma-ray spectra of BL Lac objects, comparing with the observed spectra, the EBL density can be derived \cite[e.g.][]{ ste1993,dewk1994,sta1998,sch2005,maz2007,fin2009,yang2010,dom2013,sin2014}. Several studies have been carried out to limit the EBL density by VHE observations. \cite{yang2010} and \cite{fin2009} used an assumption that the intrinsic photon index may abide by the certain values (e.g.,$\Gamma \sim$ 1.5). However, their results relay on the origins of the observed peak between $Fermi$ spectra and VHE. Various studies have been carried out to constrain the EBL density by assuming an intrinsic $\gamma$-ray spectrum \cite[e.g.][]{maz2007,ack2012,des2019,fra2019}, such as a simple power law $\mathrm{d} N / \mathrm{d} E \propto E^{-\mathrm{\Gamma}}$, as well as scaling existed EBL models \cite[e.g.][]{ste2006,kne2004,gil2009,fin2010} in amplitude.  Note that above studies, the shape of the intrinsic spectrum is still under debated, and the fitting data relay on the GeV~-~TeV band, which always are poorly observed. Furthermore, by simply scaling a factor to an EBL model, we may miss the fact that the discrepancy of the optical depth in each band is various.

The observed flux $f_{\rm{obs}}$  in the energy $E_{\rm{\gamma}}$ is given by $f_{\rm{obs}}(E_{\rm{\gamma}})=f_{\rm{int}}(E_{\rm{\gamma}}) e^{- \rm{\tau}{(E_{\rm{\gamma}},z)}}$,
where $f_{\rm{int}}$ is the intrinsic flux, and $\rm{\tau}{(E_{\rm{\gamma}},z)}$ is the optical depth of $E_\gamma$ photon which depends on the choice of the EBL template. In this paper, a model-dependent equation is introduced to describe $\tau(E_{\rm{\gamma}},z)$ as
\begin{equation}\label{tt}
\rm{log_{10}}(\tau(E_{\rm{\gamma}},z))=\rm{log_{10}}(\tau_0(E_{\rm{\gamma}},z))+\rm{a_0log_{10}}(\frac{E_{\gamma}}{E_0}),
\end{equation}
where the $\rm{a_0}$ and $\rm{E_0}$ are free parameters, and the EBL template, $\tau_0(E_{\rm{\gamma}},z)$, is given by \cite{fin2010} , in which no correlation of the VHE index with the redshift is assumed. Unlike the Polynomial fitting, this formula has a few parameters and gives the better precision of EBL approximation.  

In this paper, we employ the one-zone synchrotron self-Compton (SSC) model to estimate the intrinsic spectrum, which can be well used to produce the SEDs of BL Lacs \citep{Ghi2010b,zhang2012} for the TeV BL Lacs suffer less contamination of the emission from the accretion disk and external photon fields. In addition, we also exclude the possible radiation mechanism like the hadronic model for the consistency among objects and will take into account in the future study. Furthermore, in several cases, e.g., Mrk 501 in 1997 \citep{aha1999}, the SED of BL Lacs, if not observed at the same time, can't be explained with a single SSC component. For the reasons above, to guarantee the one-zone SSC method is appropriated, we only use quasi-simultaneous multiwavelength data to fit the intrinsic spectrum. 

Noted that the electron energy distributions (EEDs), encoding the important information about the acceleration and cooling processes of electrons in the jet, is still uncertain, and will significantly affect the shape of SEDs in the SSC scenario \citep{yan2013,qin2018a,qin2018b}. It will remind us to test kinds of EEDs in our paper   so that the best fitting result can be obtained. Commonly, three types of EEDs, i.e. a power-law EED with exponential cut-off (PLC), a power-law log-parabola (PLLP) EED and a broken power-law (BPL) EED  can be found in the literature, e.g.,\cite{zhu2016,qin2018a}. However the PLC EED does not cover SEDs very well and has been ruled out in the TeV BL Lacs scenario (please see our previous paper \cite{qin2018a}). In this paper, two types of EEDs, PLLP and BPL have been carried to fit the intrinsic SEDs of BL Lacs. The detailed description of those modes will be gave in the next section.

Throughout this work, we take the cosmological parameters, $H_0$ = 70 km s$^{-1}~$Mpc$^{-1}$, $\Omega_{\rm M}$=0.3, and $\Omega_{\rm {\Lambda}}$=0.7 to calculate the luminosity distance.

\section{MODEL AND STRATEGY }
The EBL photons of proper number density $n (\epsilon, z)$  with
energy $\epsilon$ at redshift $z$, can attenuate the observed VHE $\gamma$-rays with $E_\gamma$ by
\begin{equation}\label{tau}
\rm{\tau}{(E_{\rm{\gamma}},z)}  = c \pi r_e^2 (\frac{m^2 c^4}{E_{\gamma}})^2
\int_0^{z} dz {dt \over dz }
 \int_{\frac{ m^2c^4}{E_{\rm{\gamma}}  (1+z)}}^\infty d\epsilon \cdot
 \epsilon^{-2} n_(\epsilon, z) \bar{\varphi}[s_0(\epsilon)],
\end{equation}
where $r_e$ is the classical electron
radius, $s_0=\epsilon E_{\rm{\gamma}}/m^2c^4$, $\bar{\varphi}[s_0(\epsilon)]$ is a function given by \cite{gou1967}, and $\frac{dt}{dz}$ is the differential time of redshift.

For TeV BL Lacs with the low redshift, the $dt/dz$ can be roughly taken as $H_0^{-1}$. Usually the monochromatic approximation is used, the EBL photon absorbing gamma-ray photon with $\epsilon_{1}$ has the energy of $\epsilon_{*}^{\prime}\approx 2 / \epsilon_{1}$, where the pair-production cross section is the largest. Using the Dirac delta function, the upper limit on the EBL number density,  $n (\epsilon, z)$, can be expressed as
\begin{equation}\label{nu}
n(\epsilon, z \approx 0) \approx \epsilon^{\prime} n\left(\epsilon^{\prime}, z \approx 0\right) \delta\left(\epsilon-\epsilon^{\prime}\right).
\end{equation}

Taking Eq. (\ref{nu}) into Eq.(\ref{tau}), we obtain the density of EBL photons as
\begin{equation}
n\left(\epsilon^{\prime}, z \approx 0\right)=\frac{2 H_{0} \tau_{\gamma \gamma} E_{\gamma}}{c z_{e} \pi \rm{r_{e}^{2}} m^{2} c^{4} \overline{\varphi}(2)},
\end{equation}
where $\overline{\varphi}(2) \approx 1.787$, and $\tau_{\gamma \gamma}$ is given by fitting the observed SEDs of TeV BL Lacs with the optical depth Eq.(\ref{tt}). The error of the EBL number density is given by
\begin{equation}
\sigma(n)=\frac{2 H_{0} \sigma\left(\tau_{\gamma \gamma}\right) E_{\gamma}}{c z_{e} \pi \rm{r_{e}^{2}} m^{2} c^{4} \overline{\varphi}(2)},
\end{equation}
and $\sigma(\tau_{\gamma \gamma})$ can be roughly estimated by
 $\sigma(\tau_{\gamma \gamma})=\sigma(f_{\rm{obs}}(E_{\rm{\gamma}}))/f_{\rm{obs}}(E_{\rm{\gamma}})$.

Finally, the EBL density can be expressed as
\begin{equation}\label{ivh}
\nu I_{\nu}(z)=\frac{c}{4 \pi} \epsilon^{2} n(\epsilon, z).
\end{equation}

We assume that the emitting region of TeV BL Lacs is a sphere of the size $R$, where the magnetic field strength and a Doppler factor are $B$ and $\delta_{\rm D}$, respectively. We then fit the quasi-simultaneous SEDs data of each source with a SSC model given by \cite{fin2008}.  Furthermore, we employ the MCMC algorithm to do the SEDs fit, which is very efficient for the minimization in high-dimensional parameter space \citep{lew2002,mac2003,yan2013,ino2016,qin2018a,qin2018b}. After fitting process, we can obtain two types of probability distributions, namely, the maximum probability, which is consisted with the best-fit one obtained by minimizing the likelihood, and the marginalized probability distributions that can reflect the confident levels (C.L.) of the parameters. If two types of distributions have a similar shape and interval, then we can assert that the parameters have been constrained well.

The PLLP EED, produced by stochastic acceleration when the acceleration dominates the radiative cooling \citep{bec2006,tra2011}, is given by
\begin{equation}\label{PLLP}
 N^{\prime}(\gamma^{\prime})=K'_{\rm e}\left\{
 \begin{array}{ll}
\left(\frac{\gamma^{\prime}}{\gamma^{\prime}_{\rm c}}\right)^{-s} & \gamma^{\prime}_{\rm min}\leq \gamma^{\prime}\leq\gamma^{\prime}_{\rm c} \\
\left(\frac{\gamma^{\prime}}{\gamma^{\prime}_{\rm c}}\right)
 ^{-[s+r\log(\frac{\gamma^{\prime}}{\gamma^{\prime}_{\rm c}})]} &  \gamma^{\prime}_{\rm c}\leq\gamma^{\prime}\leq\gamma^{\prime}_{\rm
 max}\;.
 \end{array}
 \right.
\end{equation}

When no acceleration exists in the jet, the cooled EED, can be represented by the \cite{band1993}-type function. \cite{der2009} has modified this EEDs and described as the BPL shape by
\begin{eqnarray}\label{BPL}
N_{\rm e}^{\prime}(\gamma^{\prime})= K'_{\rm e}H(\gamma^{\prime};\gamma_{\rm min}^{\prime},\gamma_{\rm
max}^{\prime})\{{\gamma^{\prime
-p_1}\exp(-\gamma^{\prime}/\gamma_{\rm b}^{\prime})}
\nonumber \\
\times H[(p_{\rm 2}-p_{\rm 1})\gamma_{\rm
b}^{\prime}-\gamma^{\prime}]+[(p_{\rm 2}-p_{\rm 1})\gamma_{\rm
b}^{\prime}]^{p_{\rm 2}-p_{\rm 1}}\gamma^{\prime -p_{\rm 2}}
\nonumber \\
\times \exp(p_{\rm 1}-p_{\rm 2})H[\gamma^{\prime}-(p_{\rm
2}-p_{\rm 1})\gamma_{\rm b}^{\prime}]\},
\end{eqnarray}
where $s$ is the electron energy spectral index, $r$ is the curvature term of the EED, $\gamma^{\prime}_{\rm c}$ is the high energy cut-off, $K'_{\rm e}$ is the normalization factor of the EED, $\gamma_{\rm min}^{\prime}$ and $\gamma_{\rm max}^{\prime}$ are the minimum and maximum energies of electrons, respectively. In the Eq. (\ref{BPL}), the $H(x;x_{1},x_{2})$ is the Heaviside function:
$H(x;x_{1},x_{2})=1$ for $x_{1}\leq x\leq x_{2}$ and
$H(x;x_{1},x_{2})=0$ everywhere else, $\gamma'_{b}$ and $\gamma'_{c}$ are the break Lorentz factor in the two EEDs, and $p_{1,2}$ is the spectral index below and above $\gamma'_{\rm {b}}$.

As shown above, our model contains the size of blob $R'_{\rm{b}}$, the magnetic field $B$, the redshift $z$, the free parameters of $\rm{a_0}$ and $\rm{E_0}$ in the optical depth, the Doppler factor $\delta_{\rm {D}}$, the minimum variability $t_{\rm {v,min}}$, and the electron spectrum. However, $\gamma'_{\rm {min}}$ is always poorly constrained by SED modeling. To get $\gamma'_{\rm {min}}$ is difficult and can not been fitted directly for the origin of radio band is uncertain, so we set $\gamma'_{\rm {min}}$ to be 40-300 used in the literature, e.g.,\cite{zhang2012} and \cite{qin2018b}, in which their $\gamma'_{\rm {min}}$ constrained with the observed SEDs via a method offered by \cite{tav2010}. However, to get better fitting SEDs, we adjust $\gamma'_{\rm {min}}$ in a small range and compare with $\chi^2$/d.o.f. We get the blob's size by $R'_{\rm{b}}=\delta_{\rm {D}}ct_{\rm {v,min}}/(1+z)$. According to \citep{xie2001,fos2008}, the timescales of variability from hours to intra-day have been observed in the BL Lacs, therefore, we simply set $t_{\rm v,min}$ to be 24 hours for the source without minimum variability. Our model is not sensitive to $\gamma'_{\rm {max}}$ for the Klein-Nishina (KN) effect that can modify the SSC spectra at high energy\citep{mod2005}, so we roughly set $\gamma'_{\rm {max}}$ as 100$\times\gamma'_b$ ($\gamma'_c$) \citep{zhang2012}.

\section{APPLICATION AND RESULTS}

In order to study EBL by the pair production process, six TeV BL Lace objects with (quasi-) simultaneous multi-waveband SEDs are considered. The sample contains one intermediate synchrotron peaked BL Lac object (IBL, $10^{14}$Hz $<~\nu_{\rm{pk}}~<10^{15}$Hz), 3C~66A, and six high synchrotron peaked BL Lac objects
(HBL, $10^{15}$Hz $>~\nu_{\rm{pk}}$), namely, 1E~S1218+30.4, Mrk~421, Mrk~501, PKS0447-439, W~com (with high and low stage). Their (quasi-) simultaneous multi-waveband data, $\gamma'_{\rm {min}}$ and $t_{\rm v,min}$ are complied by \cite{zhang2012} except for PKS 0447-439, which the  data is getting from \cite{pra2012}.

The EBL spectrum presented in Figure 1-7 is estimated directly from fitting observed SEDs of TeV BL Lasc objects with one-zone SSC model, and the model parameters are summarized in Table \ref{Table:1}-\ref{Table:2}. We also plot the reduced EBL density in the the two panels of Fig.8 based on the value of EBL density. Here we briefly describe the strategy for fitting seven BL Lacs and interpret their result.

\begin{table*}
\centering
%%\small
\caption{The model parameters of the best-fits and the marginalized 68\%
confident intervals (CI) are listed for BPL EEDs. \label{Table:1}}
\resizebox{\textwidth}{!}{
%\begin{threeparttable}
\begin{tabular}{lccc ccccc cc}
\hline \hline
Source name & $z$ & $ B~(0.1G)$ &Log$_{10}[{\gamma'_{\rm b}}]$ & $\delta_{\rm D}~(10)$& Log$_{10}[K'_{\rm e}]$ & $p_1$ & $p_2$ &$a_0$&$E_0$& $\chi^2/{\rm {d.o.f}}$\\
~[1] & [2] & [3] & [4] & [5] & [6] &[7] & [8] & [9]&[10]&[11] \\
\hline\hline
 1ES1218+30.4& 0.18& 4.68& 5.18& 1.06&54.27& 2.39& 6.18& 0.70& 7.25& 0.51\\
      (68\%~CI)&~-~& 1.97~-~ 4.98& 5.07~-~ 5.30& 1.00~-~ 1.53&53.71~-~54.51& 2.25~-~ 2.44& 4.00~-~ 7.00& 0.48~-~ 3.27& 1.90~-~ 8.40&~-~\\
\hline
         3C66A& 0.44& 0.97& 4.14& 3.25&50.80& 1.19& 5.03& 0.46& 5.23&1.23\\
      (68\%~CI)&~-~& 0.99~-~ 1.47& 4.07~-~ 4.16& 2.80~-~ 3.28&50.33~-~51.41& 1.07~-~ 1.36& 5.00~-~ 5.11& 0.62~-~ 3.21& 0.50~-~ 8.41&~-~\\
\hline
             MRK421& 0.03& 0.46& 5.12& 4.19&52.02& 2.07& 4.87& 7.02& 7.48&1.85\\
      (68\%~CI)&~-~& 0.41~-~ 0.51& 5.10~-~ 5.14& 4.06~-~ 4.36&51.95~-~52.07& 2.05~-~ 2.09& 4.76~-~ 5.01& 3.11~-~ 8.00& 5.57~-~ 8.00&~-~\\
\hline
              MRK501& 0.03& 0.98& 5.37& 1.44&53.96& 2.38& 4.04& 3.76& 7.56& 0.37\\
      (68\%~CI)&~-~& 0.60~-~ 1.26& 5.29~-~ 5.57& 1.29~-~ 1.75&53.65~-~54.62& 2.31~-~ 2.53& 3.75~-~ 8.30& 0.86~-~ 4.00& 3.08~-~ 8.69&~-~\\
\hline
         PKS0447-439& 0.11& 0.56& 4.80& 2.24&54.89& 2.35& 4.39& 0.29& 0.18& 0.42\\
      (68\%~CI)&~-~& 0.76~-~ 1.36& 4.64~-~ 4.89& 1.65~-~ 2.05&54.41~-~55.60& 2.27~-~ 2.57& 4.35~-~ 6.24& 0.00~-~ 5.61& 1.60~-~12.16&~-~\\
\hline
         Wcom& 0.10& 0.10& 4.51& 4.10&52.96& 1.85& 3.75& 0.41& 0.28&1.67\\
      (68\%~CI)&~-~& 0.13~-~ 0.25& 4.43~-~ 4.58& 3.11~-~ 3.87&52.97~-~53.42& 1.85~-~ 2.01& 3.71~-~ 3.92& 0.00~-~ 1.44& 0.61~-~ 5.27&~-~\\
\hline
         Wcom flare& 0.10& 0.10& 4.32& 3.58&52.46& 1.62& 3.49& 0.31& 0.22&0.86\\
      (68\%~CI)&~-~& 0.10~-~ 0.93& 4.27~-~ 4.43& 1.55~-~ 3.00&52.78~-~53.79& 1.73~-~ 2.03& 3.52~-~ 3.76& 0.00~-~ 0.79& 0.10~-~ 5.37&~-~\\
\hline
 \end{tabular}
%\end{threeparttable}
}

%\vskip 0.4 true cm
\end{table*}

\begin{table*}
\centering
%%\small
\caption{The model parameters of the best-fits and the marginalized 68\%
confident intervals (CI) are listed for PLLP EEDs. \label{Table:2}}
\resizebox{\textwidth}{!}{
%\begin{threeparttable}
\begin{tabular}{lccc ccccc cc}
\hline \hline
Source name & $z$ & $ B~(0.1G)$ &Log$_{10}[{\gamma'_{\rm c}}]$ & $\delta_{\rm D}~(10)$& Log$_{10}[K'_{\rm e}]$ & $s$ & $r$ &$a_0$&$E_0$& $\chi^2/{\rm {d.o.f}}$\\
~[1] & [2] & [3] & [4] & [5] & [6] &[7] & [8] & [9]&[10]&[11] \\
\hline\hline
   1ES1218+30.4& 0.18& 0.55& 5.54& 2.19&41.05& 2.65& 8.55& 0.20& 3.15& 0.35\\
      (68\%~CI)&~-~& 0.25~-~ 3.42& 5.19~-~ 5.54& 1.00~-~ 2.19&41.02~-~41.70& 2.61~-~ 2.69& 3.82~-~14.99& 0.72~-~ 3.38& 2.41~-~ 8.64&~-~\\
\hline
          3C66A& 0.44& 1.02& 3.68& 3.13&46.43& 1.67& 1.18& 0.81& 0.85&1.28\\
      (68\%~CI)&~-~& 0.92~-~ 1.32& 3.58~-~ 3.74& 2.79~-~ 3.27&46.28~-~46.64& 1.57~-~ 1.74& 1.11~-~ 1.26& 0.91~-~ 4.00& 0.11~-~ 8.18&~-~\\
\hline
         MRK421& 0.03& 0.47& 4.61& 4.16&42.37& 2.12& 1.18& 1.32& 4.54&1.99\\
      (68\%~CI)&~-~& 0.44~-~ 0.55& 4.57~-~ 4.66& 3.95~-~ 4.29&42.25~-~42.45& 2.10~-~ 2.15& 1.13~-~ 1.26& 0.85~-~ 3.38& 0.10~-~ 4.15&~-~\\
\hline
         MRK501& 0.03& 0.70& 4.27& 1.70&43.79& 2.16& 0.57& 1.27& 6.48& 0.32\\
      (68\%~CI)&~-~& 0.36~-~ 1.00& 4.29~-~ 4.85& 1.44~-~ 2.13&42.36~-~43.77& 2.18~-~ 2.52& 0.56~-~ 0.78& 1.75~-~ 8.26& 2.51~-~14.99&~-~\\
\hline
    PKS0447-439& 0.11& 0.69& 4.11& 2.07&45.10& 2.43& 0.80& 0.70& 6.06& 0.38\\
      (68\%~CI)&~-~& 0.52~-~ 0.87& 3.78~-~ 4.29& 1.89~-~ 2.34&44.60~-~45.95& 2.31~-~ 2.55& 0.58~-~ 0.98& 0.01~-~ 8.36& 2.30~-~ 9.99&~-~\\
\hline
         Wcom& 0.10& 0.24& 3.91& 3.14&45.44& 2.09& 0.70& 0.43&12.60&2.05\\
      (68\%~CI)&~-~& 0.22~-~ 0.38& 3.85~-~ 3.97& 2.72~-~ 3.23&45.28~-~45.57& 2.07~-~ 2.17& 0.66~-~ 0.75& 0.00~-~ 8.38& 3.12~-~12.71&~-~\\
\hline
         Wcom flare& 0.10& 1.54& 3.59& 1.18&46.53& 2.16& 0.60& 5.80& 0.46&1.03\\
      (68\%~CI)&~-~& 1.40~-~ 1.71& 3.56~-~ 3.71& 1.12~-~ 1.24&46.23~-~46.62& 2.14~-~ 2.20& 0.59~-~ 0.68& 2.37~-~ 9.99& 0.10~-~12.36&~-~\\
\hline
\end{tabular}
%\end{threeparttable}
}
\end{table*}
%the fitting figure

1.\emph{1ES1218+30.4}. As an HBL, we take $t_{\rm {v,min}}$=12 hr from the light curve \citep{acc2010}. \cite{zhang2012} used $\gamma'_{\rm {min}}$=300. However, in our paper, we set $\gamma'_{\rm {min}}=100$ to get a better fitting result. From Fig. 1, we can see that the SED is well fitted by BPL and PLLP EEDs. The reduced EBL density shown in the right panel of Fig.8 is lower than that obtained by galaxy counts in the BPL EEDs, but it is roughly consistent with Stecker06-baseline \citep{ste2006} in the PLLP EEDs shown in the left panel of Fig.8. It is noted that the lower errors are not plotted for their larger uncertain. Besides, the 1$\sigma$ contour shows that the distribution of $a_0$ and $E_0$ are not constrained very well. The proper EBL density obtained by this source could be between 1.5 and 11 n W m$^{-2}$ sr$^{-1}$.

2.\emph{3C 66A}
AS an IBL, its redshift is still under debated. In this paper, we set $z=0.44$ to fit the SEDs. We take $t_{\rm {v,min}}=12hr$ and $\gamma'_{\rm {min}}=200$ \citep{zhang2012}. As shown in Fig.2, we can see that the fitting SEDs are very well in two types of EEDs. In the left panel of Fig.8, the EBL density in BPL EEDs is consistent with that of \cite{fin2010}, and in PLLP EEDs, it is higher than that of \cite{fin2010} in each band. Anyway, we can't rule out both results, which need to distinguish the SEDs via further work using the 20~-~50 KeV observation.

3.\emph{Mrk 421}
As an HBL object, we take $t_{\rm {v,min}}=3hr$ and $\gamma'_{\rm {min}}=180$ \citep{zhang2012}. As shown in Fig.3, two SEDs fail to fit the GeV-TeV spectrum, disfavoring the one-zone SSC scenario. Besides, the EBL density can't be reduced in BPL EEDs, and it is significantly lower than that obtained by galaxy counts in the PLLP EEDs, shown in the right panel of Fig.8.

4.\emph{MRK 501}
As an HBL object, we use the SEDs in the low stage compiled by \cite{zhang2012}. We take $t_{\rm {v,min}}=12hr$ and $\gamma'_{\rm {min}}=150$ \citep{zhang2012}. However $\gamma'_{\rm {min}}=250$ gives a better fitting result. As shown in Fig.4, the SEDs fitted by two types of EEDs are rational. However, like Mrk 421, the reduced EBL density by two models should be ruled out. Further research, such as the simultaneous $Fermi$ observation should be taken into account.

5.\emph{PKS 0447-439}
It is an HBL, we set $\gamma'_{\rm {min}}=100$ and $t_{\rm {v,min}}=24 hr$ to build the SEDs  \citep{qin2018b}. We can find that the modeling SEDs favor the observation both in two EEDs. However, as shown in the right panel of the Fig.8, the EBL density is lower than that given by galaxy counts in BPL EEDs and should be discarded. The EBL density obtained in the PLLP model, shown in the right panel of Fig.8, is about 6-20 n W m$^{-2}$ sr$^{-1}$ slightly higher than that givem by \cite{ste2006}.

6.\emph{W com}
As a TeV IBL object with the redshift of 0.10, we take $t_{\rm {v,min}}=12hr$ and $\gamma'_{\rm {min}}=200$ in low and flare stages. For the low stage, as shown in Fig.6, both SEDs have lost the observed data between $10^{24}-10^{25}$Hz band. In the Fig.7, like PKS 0447-439, the EBL results are valid in PLLP EEDs and about 10-20 n W m$^{-2}$ sr$^{-1}$,  which are similar to the result given by \cite{ste2006}. For the flare stage, as shown in Fig.7, the fitting SEDs seem to cover multi-waveband spectra although insufficient observation in GeV-TeV band. Unlike other sources, the reduced EBL density in BPL EEDs is more reasonable than that in PLLP EEDs, which is similar to the result given by \cite{ste2006}, and it is about 10 n W m$^{-2}$ sr$^{-1}$. We also note that the EBL density in PLLP EEDs looks very weird.

%the 1ES1218+30.4
\begin{figure*}
\centering
%\flushright
\subfigure[]{
\includegraphics[height=6.cm,width=7.5cm]{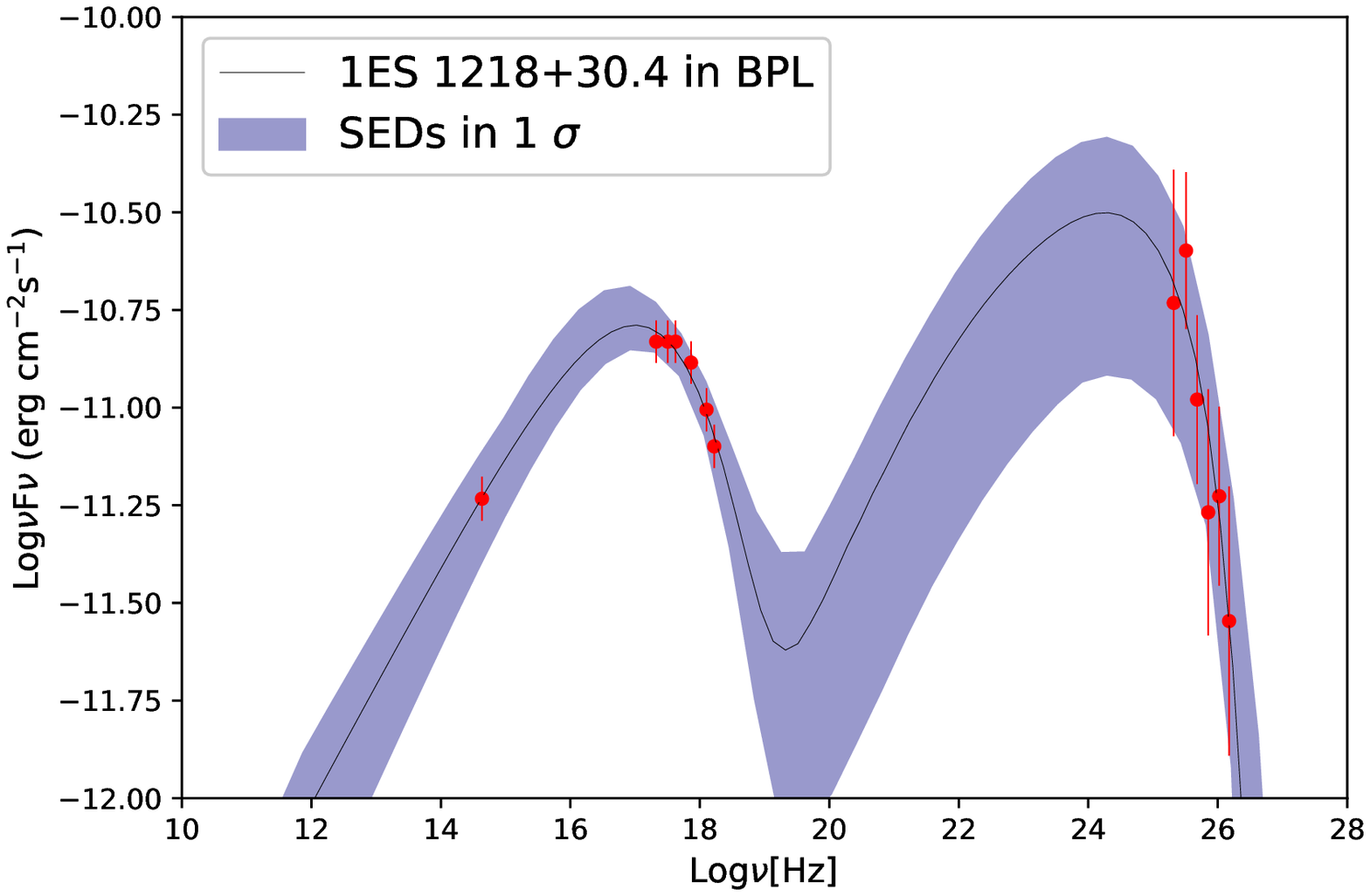}
%\caption{fig1}
}
\quad
\subfigure[]{
\includegraphics[height=6.cm,width=7.5cm]{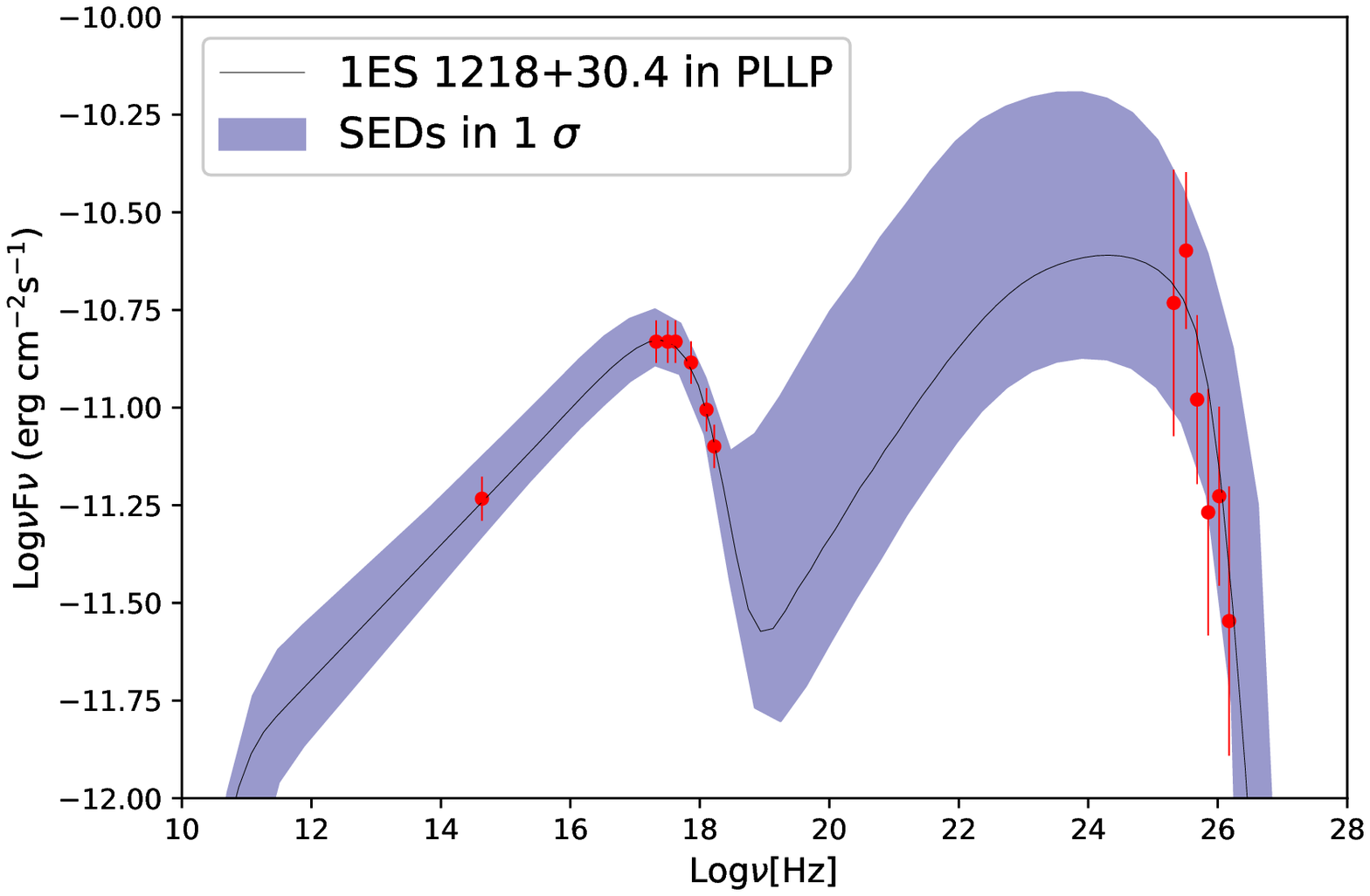}
}
%\flushright
%\centering
\quad\quad\quad\quad\quad\quad
\subfigure[]{
\includegraphics[height=5.8cm,width=7.cm]{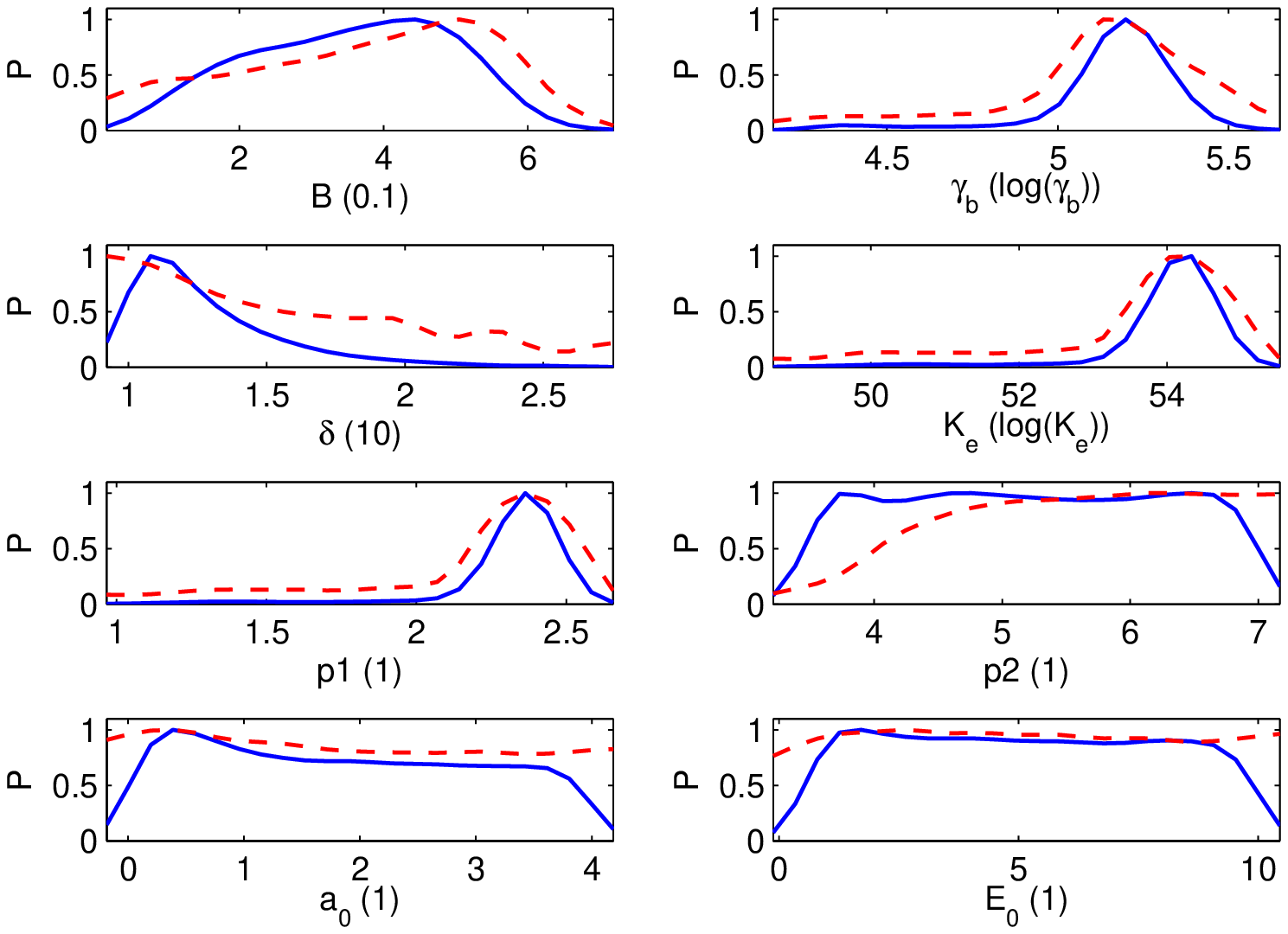}
}
\quad\quad\quad\quad
\subfigure[]{
\includegraphics[height=5.8cm,width=7.cm]{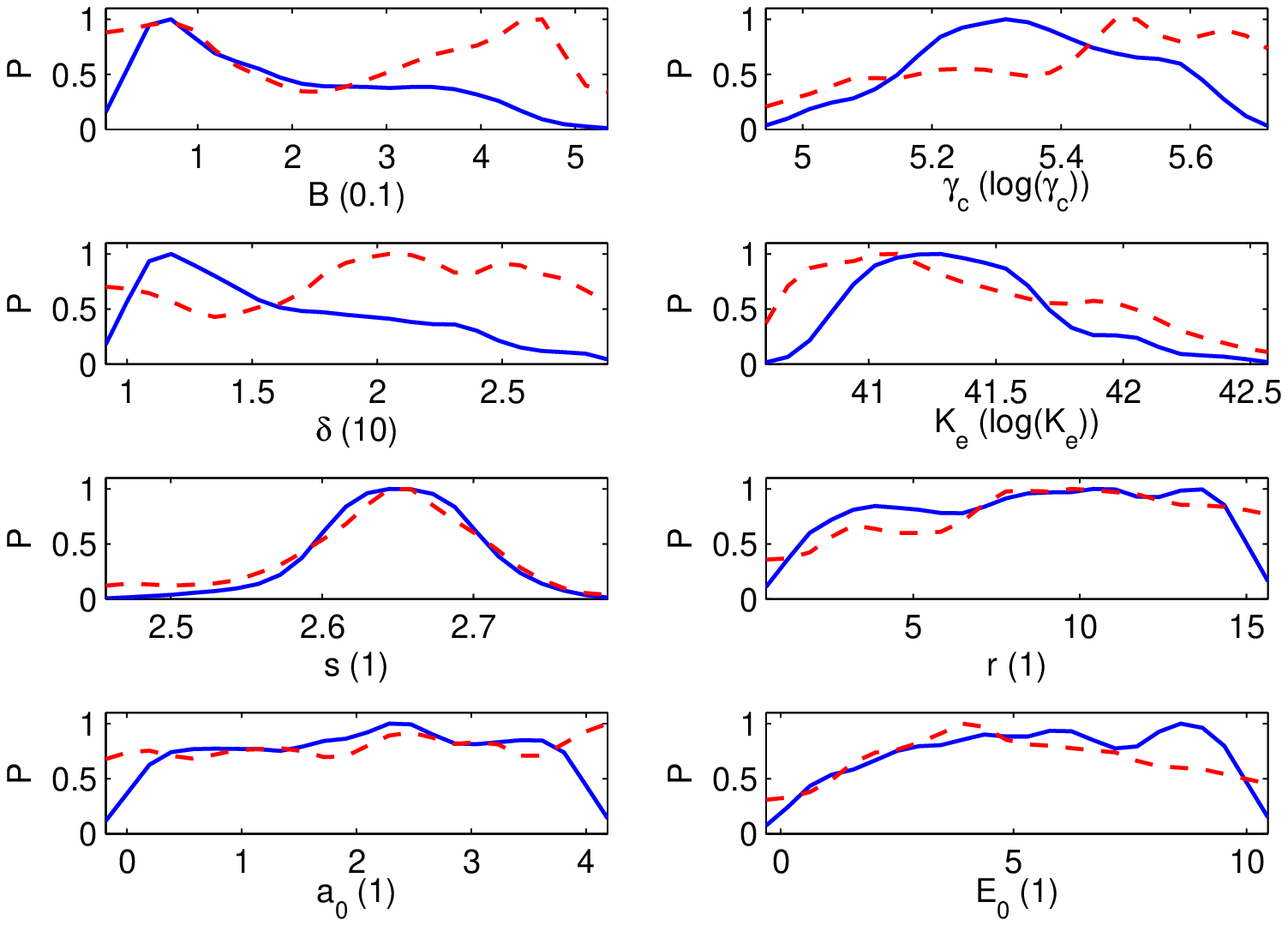}
}
\caption{The SEDs of 1ES1218+30.4 fitted by two types of EEDs. The solid black and the shaded areas in
cyan represent the best fitting SEDs and the contours under the 1$\sigma$ error bands, respectively. (a) Fitting by BPL EEDs.(b)Fitting by PLLP EEDs. (c) and (d) The distributions of parameters in BPL and PLLP EEDs,where the dotted lines show the maximum likelihood distributions, the solid lines show the marginalized probability distributions.}
\end{figure*}

\begin{figure*}
%\centering
\flushright
\subfigure[]{
\includegraphics[height=6.cm,width=7.5cm]{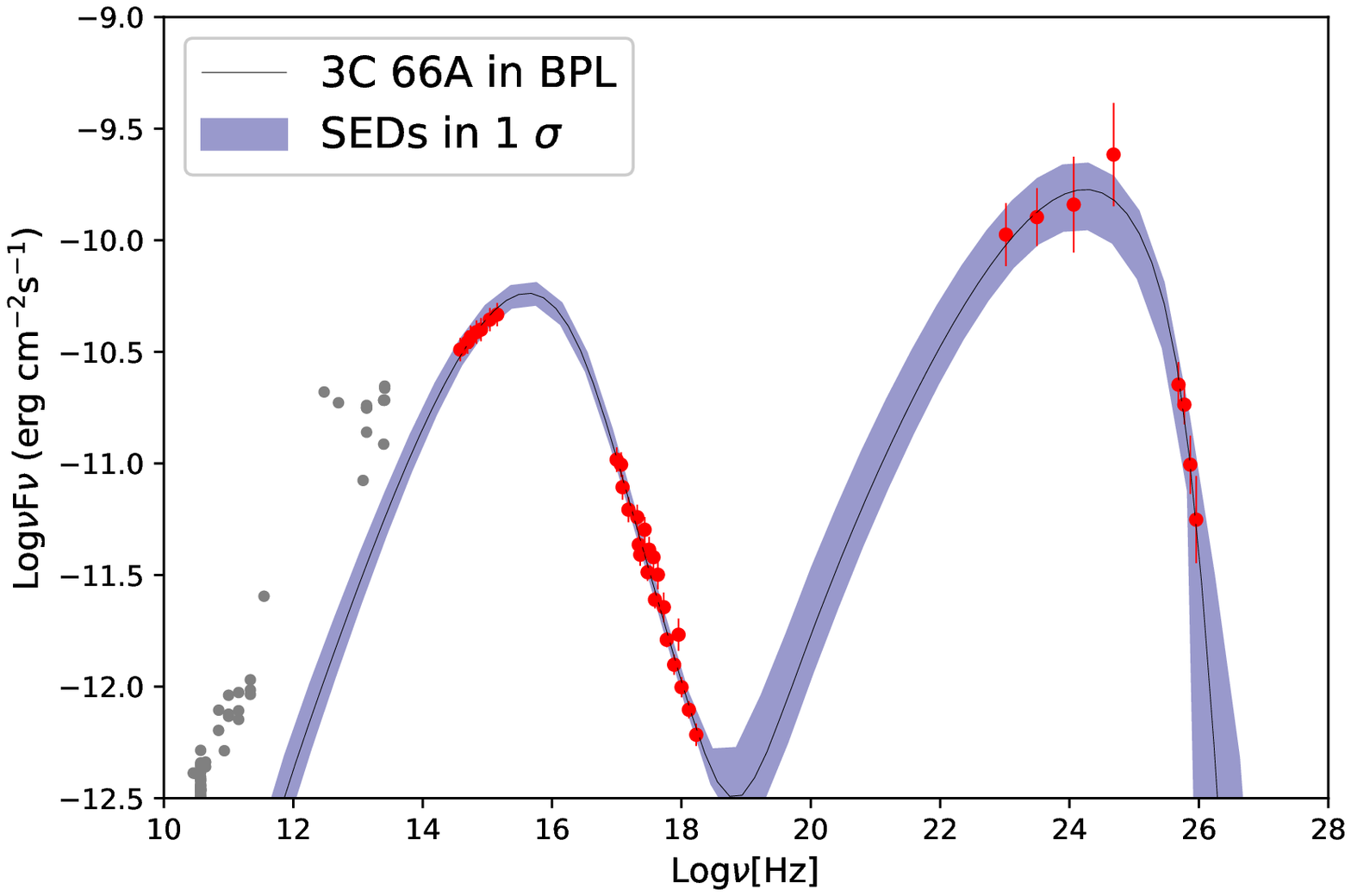}
%\caption{fig1}
}
\quad
\subfigure[]{
\includegraphics[height=6.cm,width=7.5cm]{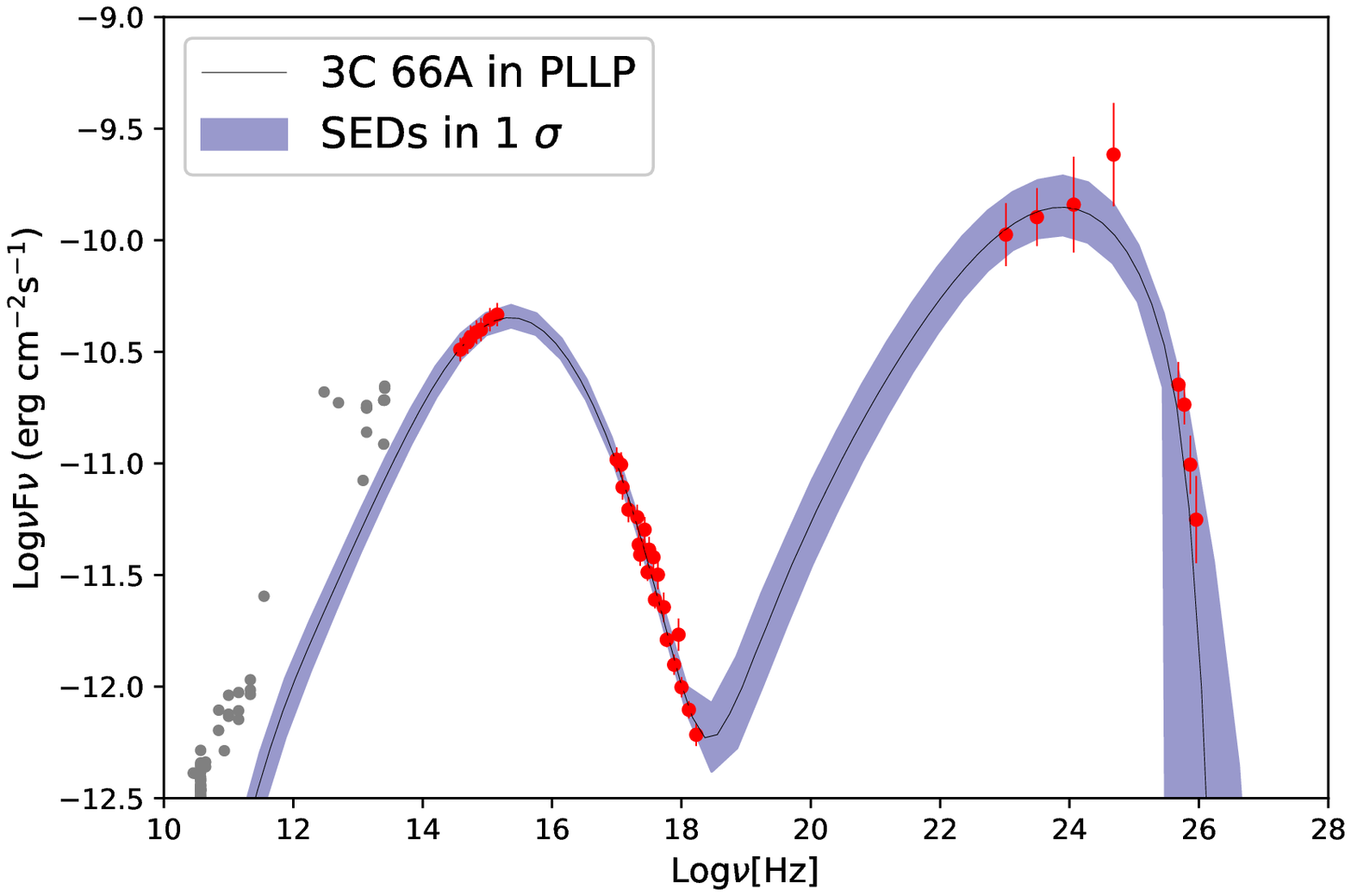}
}
\flushright
%\centering
\quad\quad\quad\quad\quad\quad
\subfigure[]{
\includegraphics[height=5.8cm,width=7.cm]{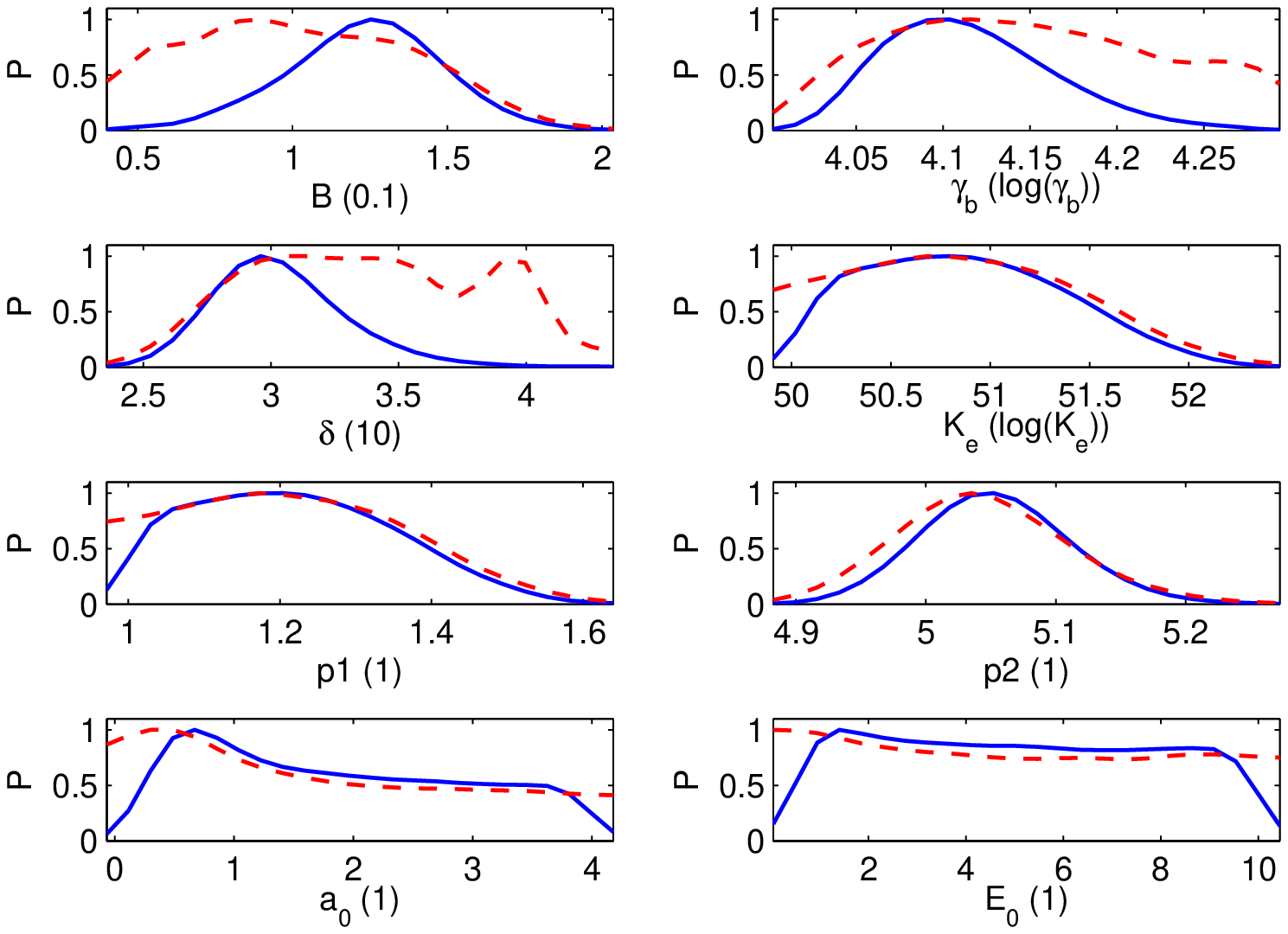}
}
\quad\quad\quad\quad
\subfigure[]{
\includegraphics[height=5.8cm,width=7.cm]{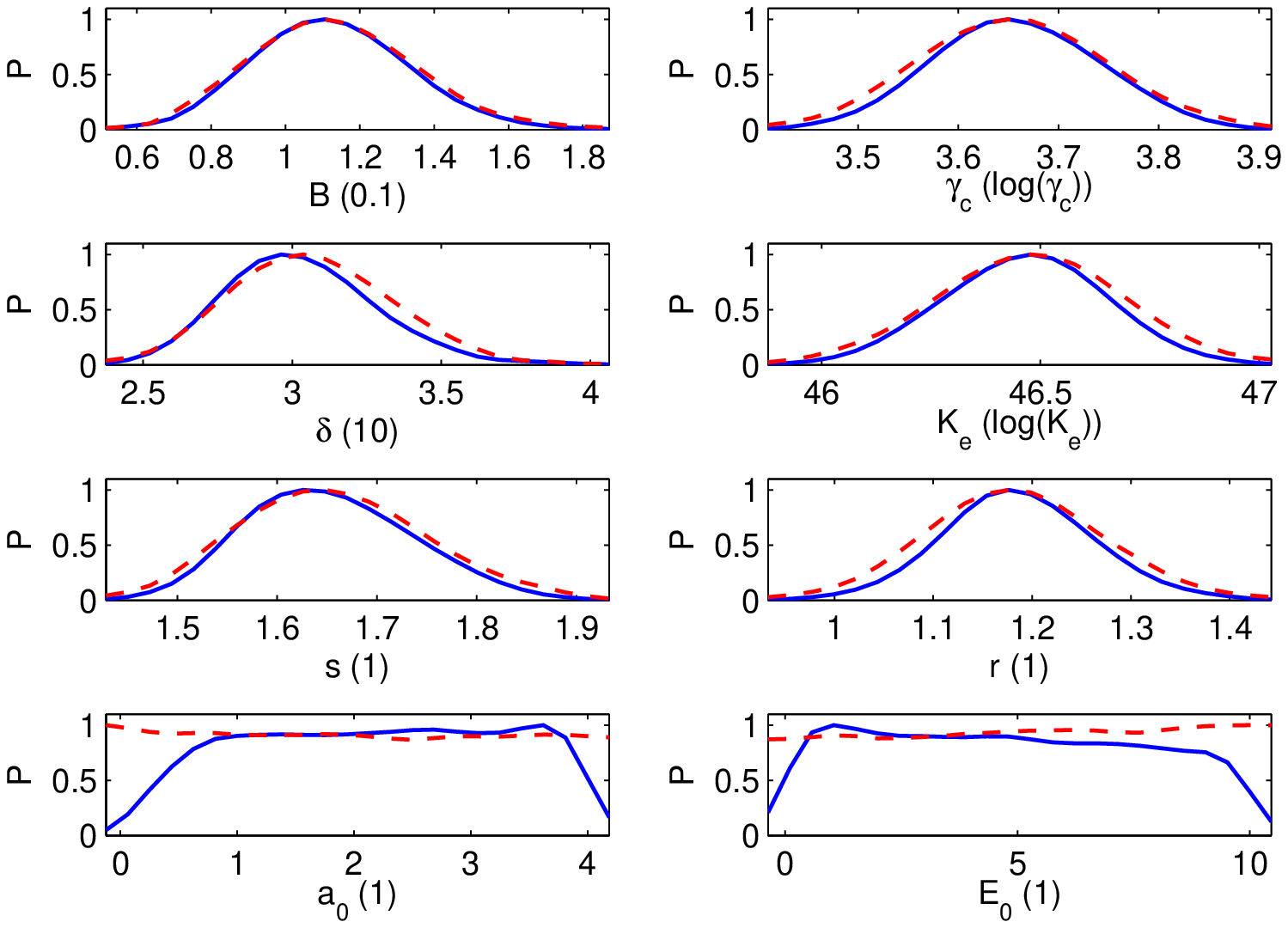}
}
\caption{As Fig. 1, but for 3C 66A.}
\end{figure*}

%+++++++++++++++++++++++++++++++++++++++++
\begin{figure*}
%\centering
\flushright
\subfigure[]{
\includegraphics[height=6.cm,width=7.5cm]{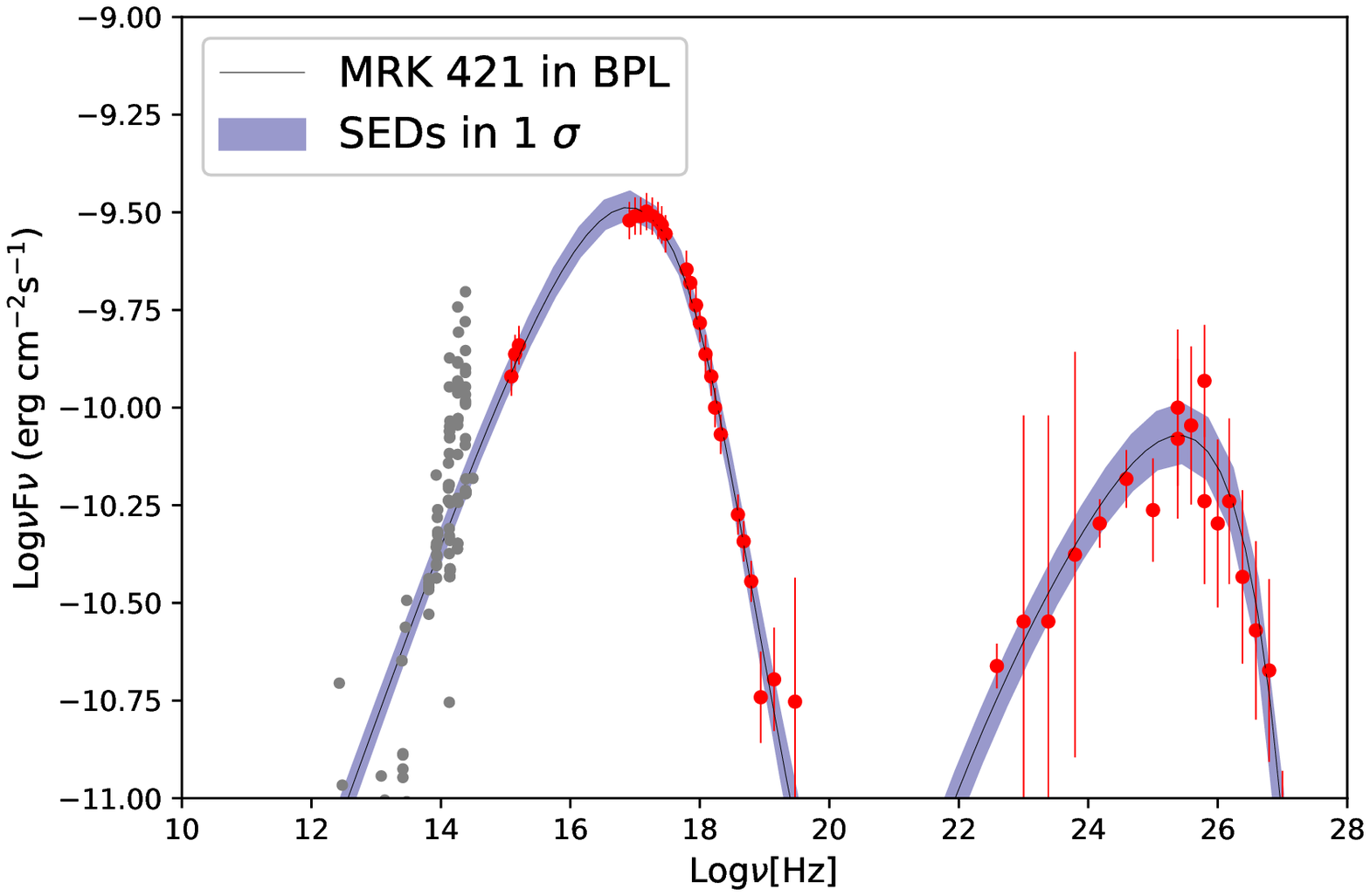}
%\caption{fig1}
}
\quad
\subfigure[]{
\includegraphics[height=6.cm,width=7.5cm]{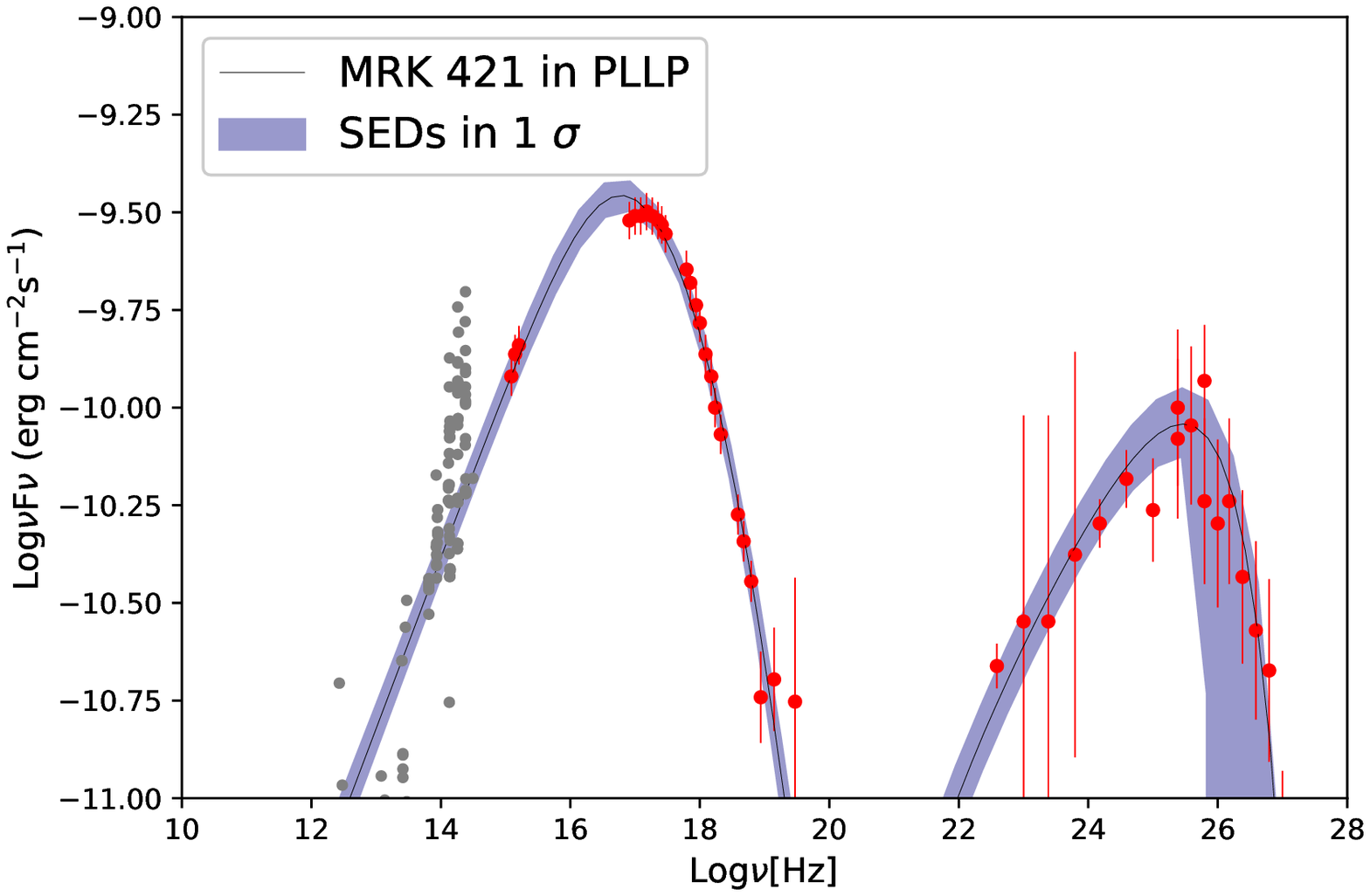}
}
\flushright
%\centering
\quad\quad\quad\quad\quad\quad
\subfigure[]{
\includegraphics[height=5.8cm,width=7.cm]{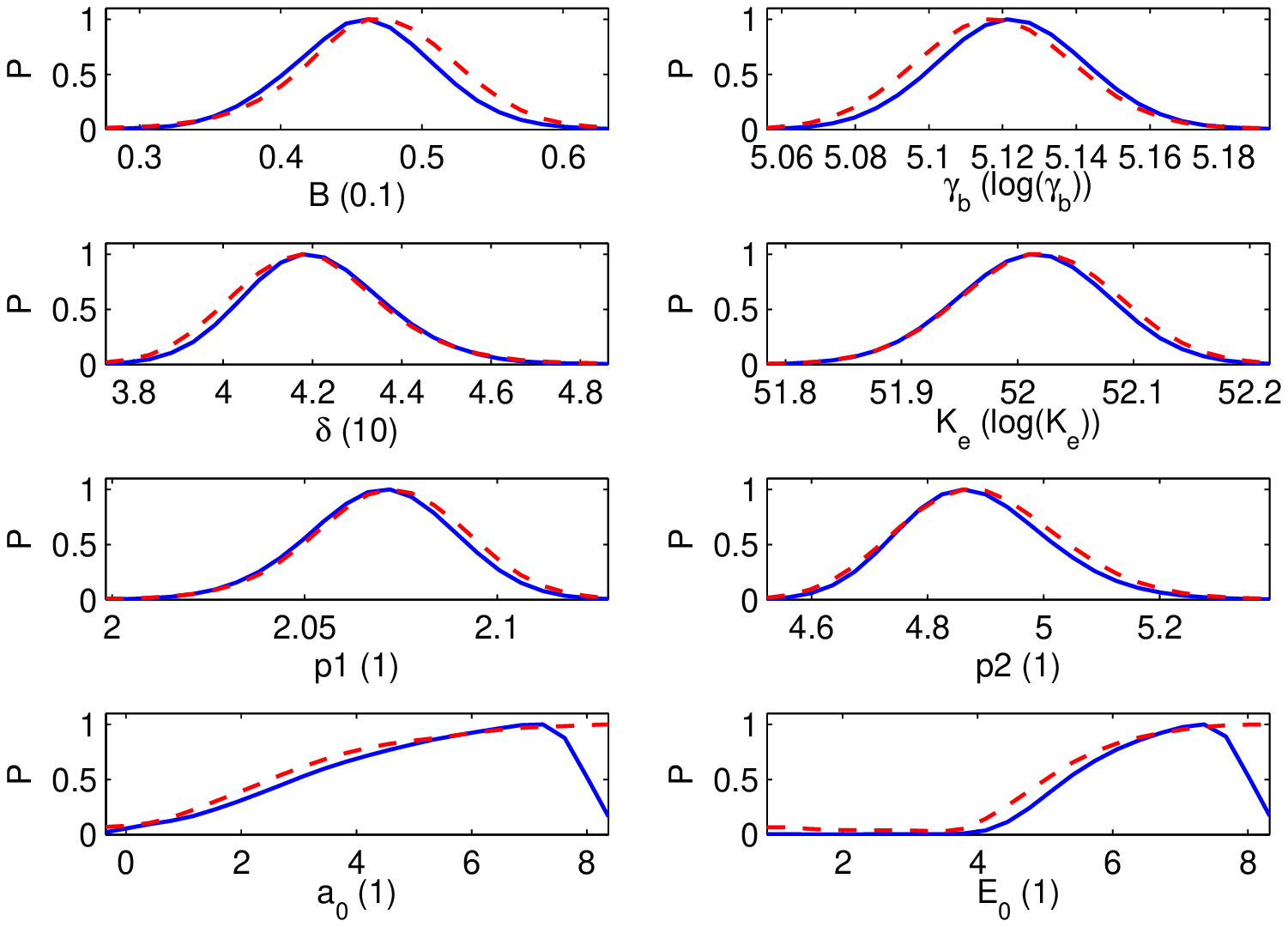}
}
\quad\quad\quad\quad
\subfigure[]{
\includegraphics[height=5.8cm,width=7.cm]{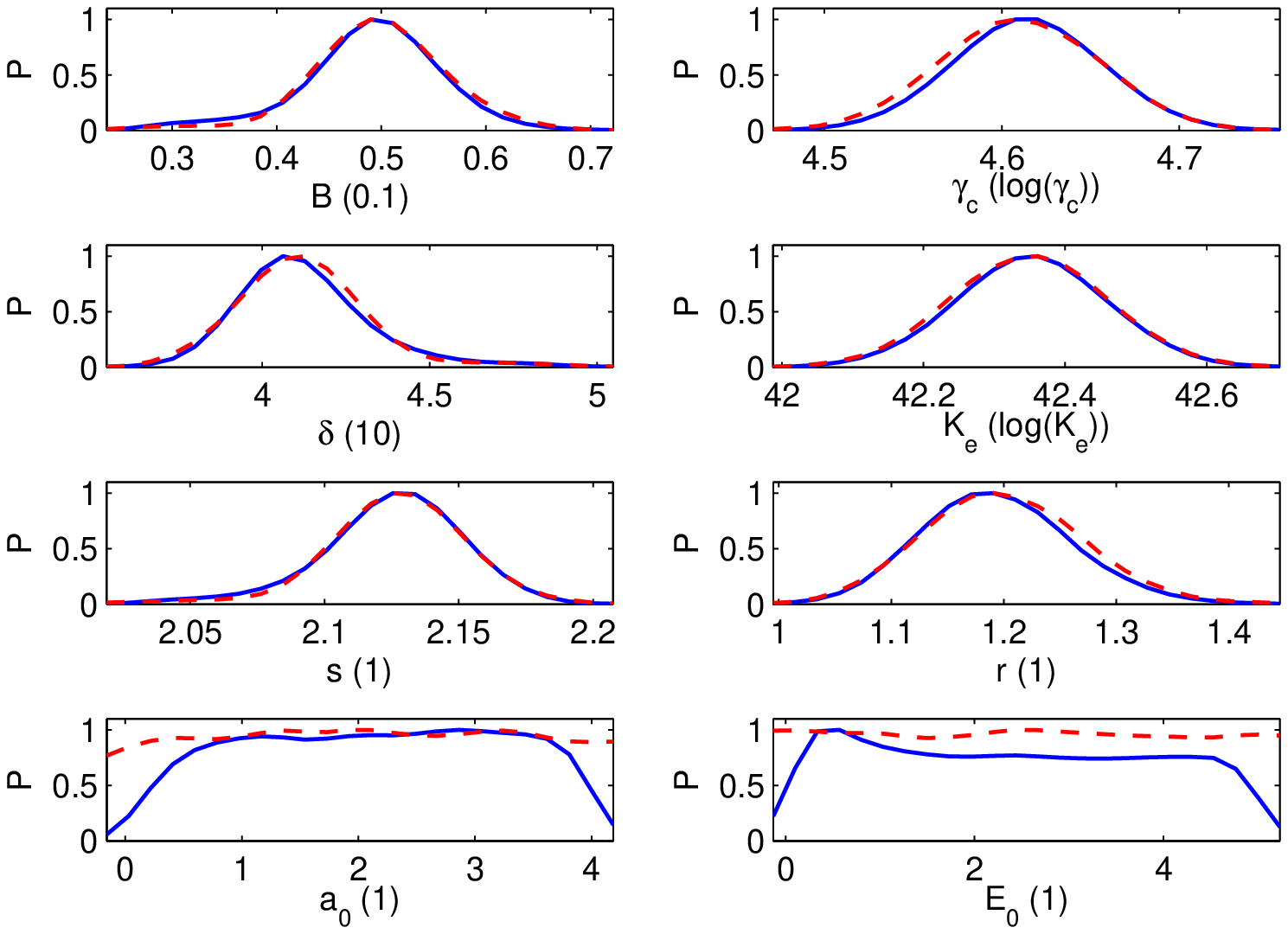}
}
\caption{As Fig. 1, but for MRK 421.}
\end{figure*}

\begin{figure*}
%\centering
\flushright
\subfigure[]{
\includegraphics[height=6.cm,width=7.5cm]{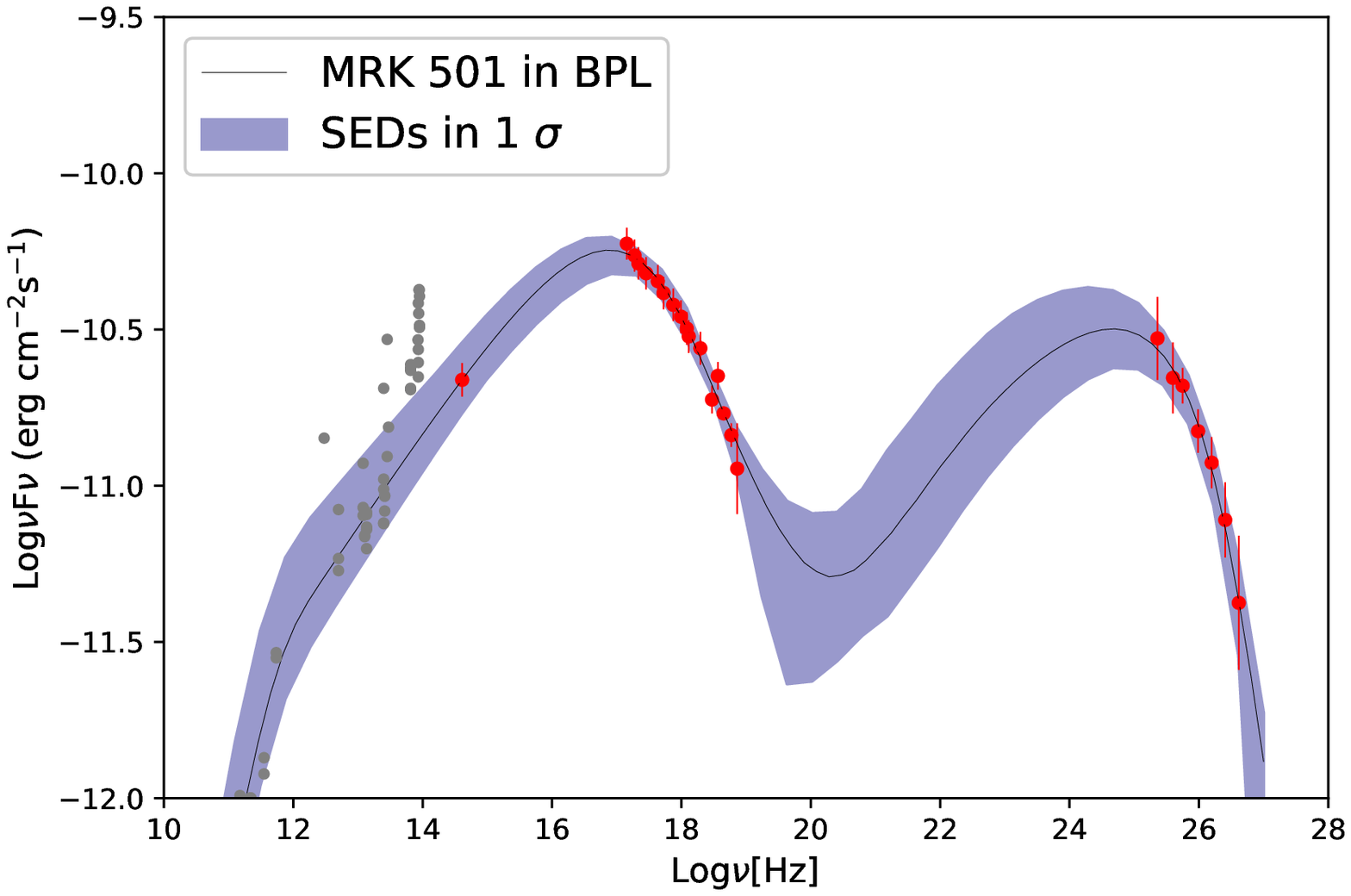}
%\caption{fig1}
}
\quad
\subfigure[]{
\includegraphics[height=6.cm,width=7.5cm]{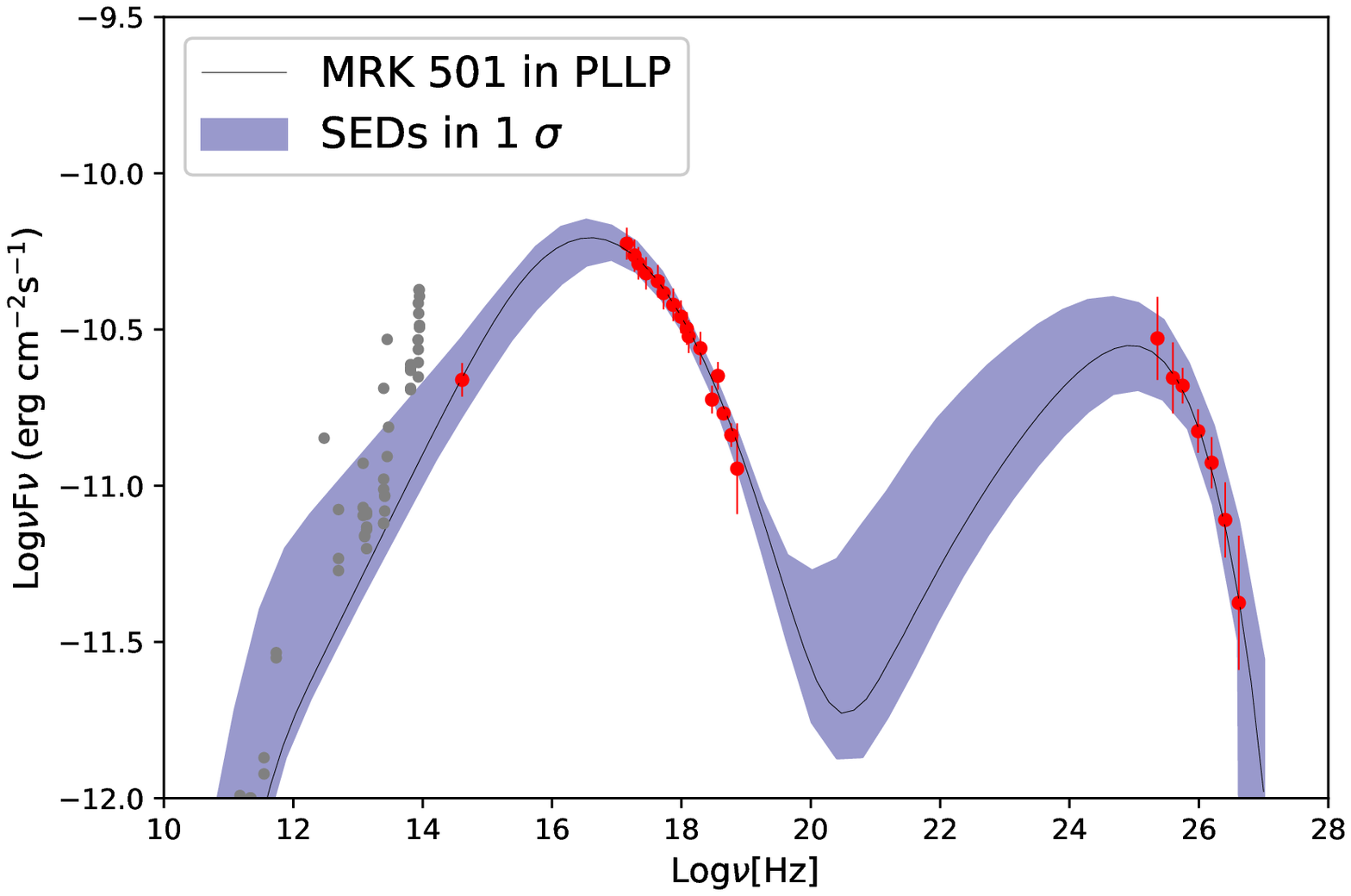}
}
\flushright
%\centering
\quad\quad\quad\quad\quad\quad
\subfigure[]{
\includegraphics[height=5.8cm,width=7.cm]{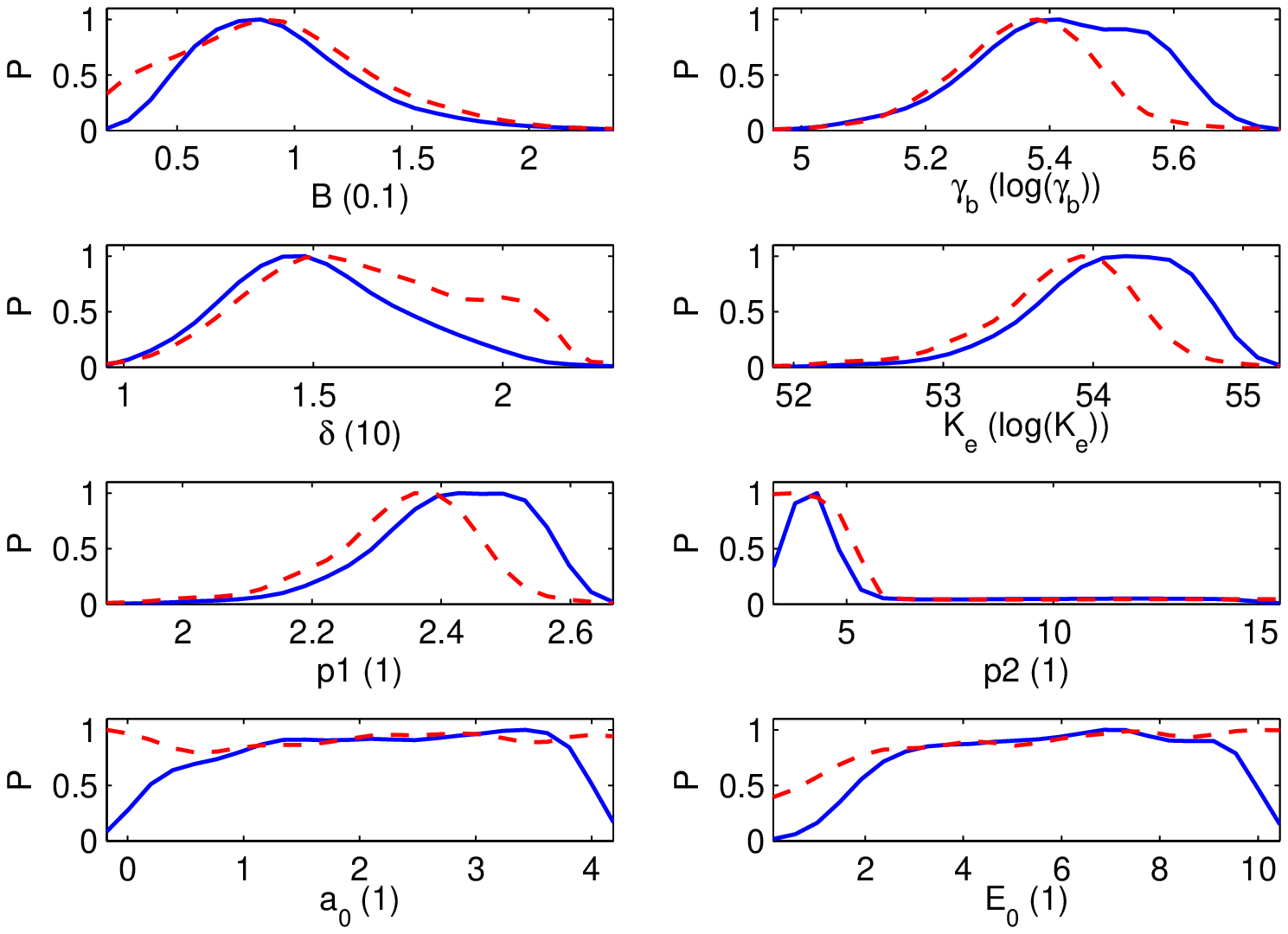}
}
\quad\quad\quad\quad
\subfigure[]{
\includegraphics[height=5.8cm,width=7.cm]{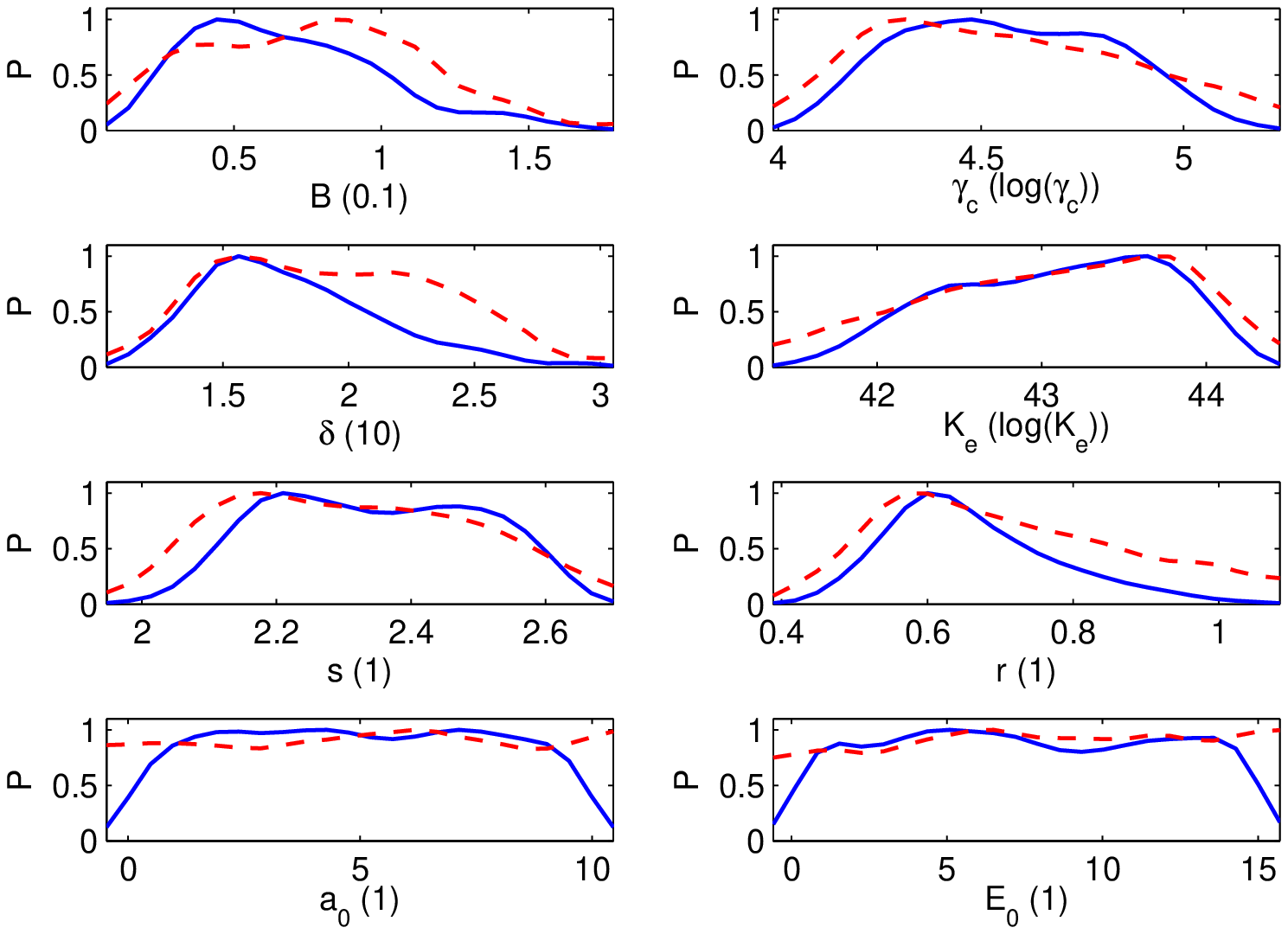}
}
\caption{As Fig. 1, but for MRK 501.}
\end{figure*}

\begin{figure*}
%\centering
\flushright
\subfigure[]{
\includegraphics[height=6.cm,width=7.5cm]{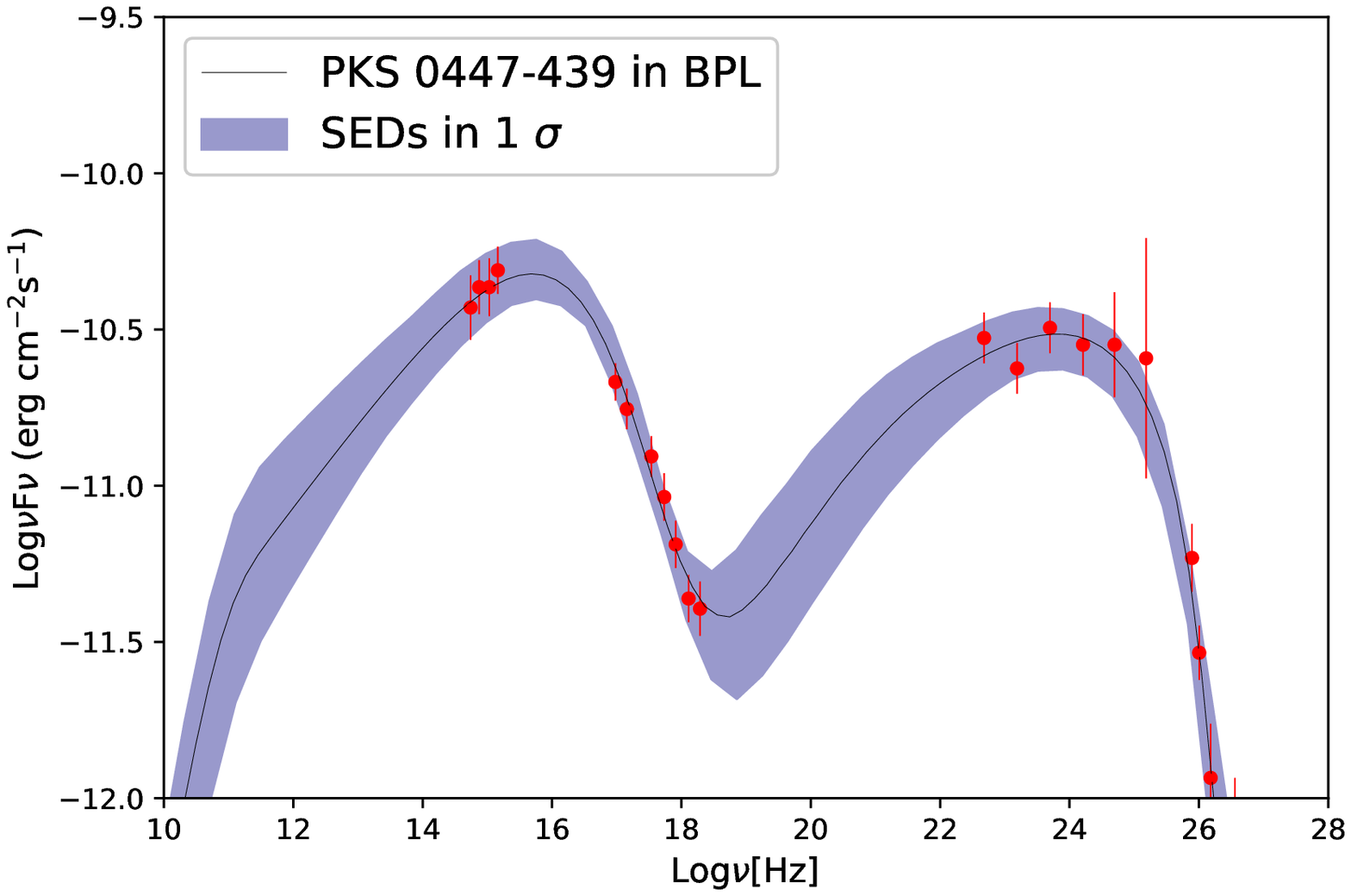}
%\caption{fig1}
}
\quad
\subfigure[]{
\includegraphics[height=6.cm,width=7.5cm]{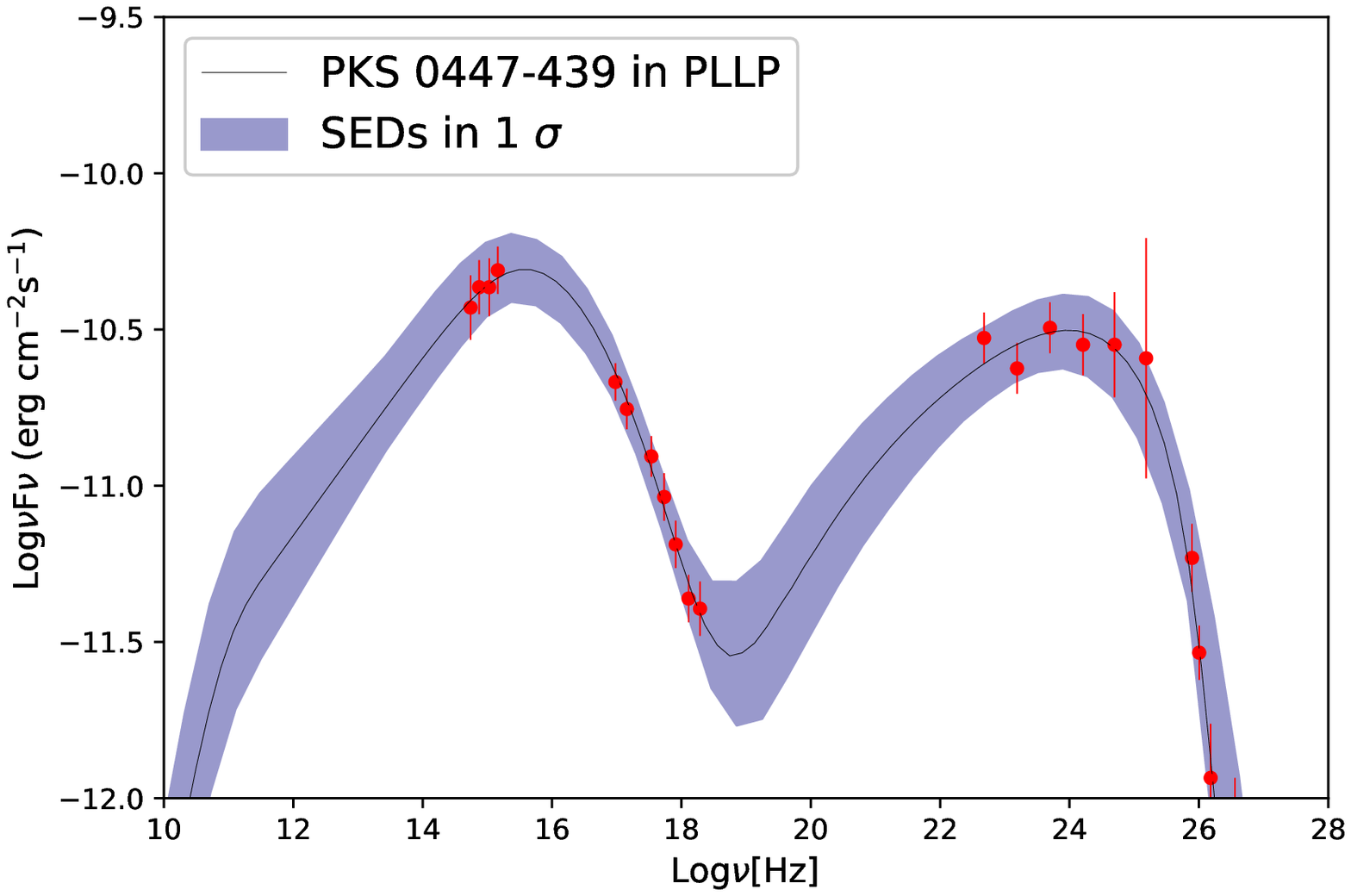}
}
\flushright
%\centering
\quad\quad\quad\quad\quad\quad
\subfigure[]{
\includegraphics[height=5.8cm,width=7.cm]{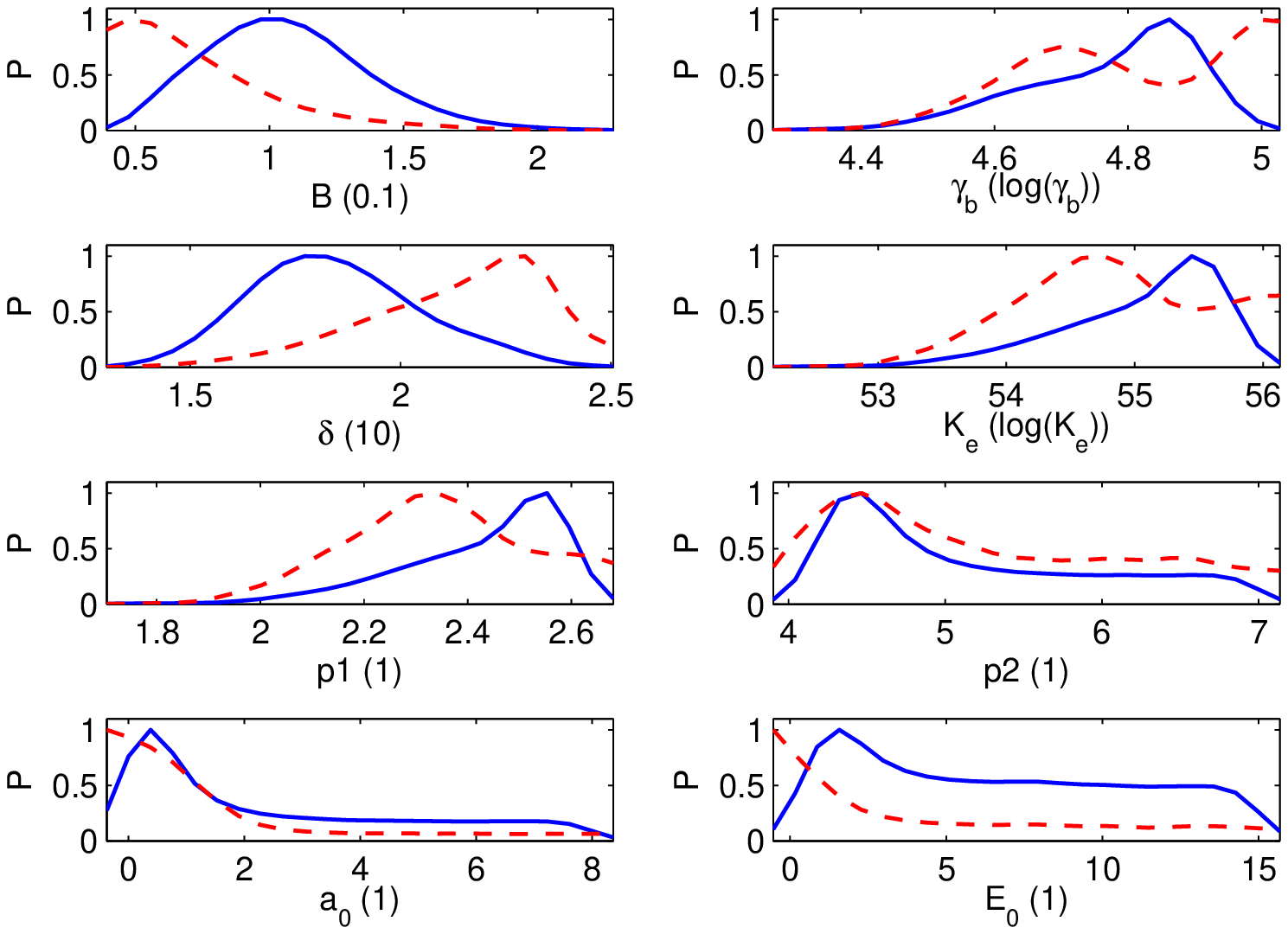}
}
\quad\quad\quad\quad
\subfigure[]{
\includegraphics[height=5.8cm,width=7.cm]{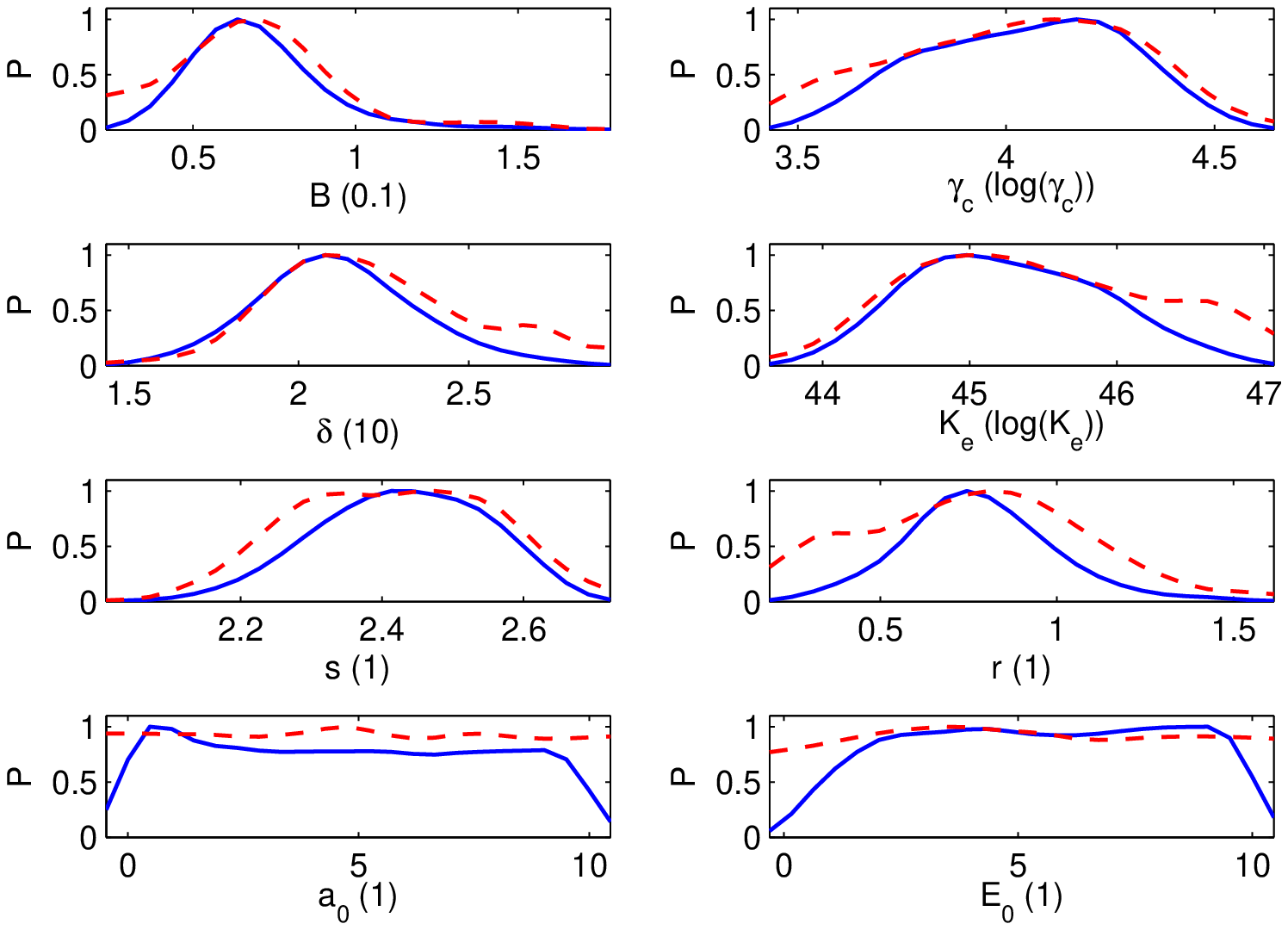}
}
\caption{As Fig. 1, but for PKS0447-439.}
\end{figure*}

\begin{figure*}
%\centering
\flushright
\subfigure[]{
\includegraphics[height=6.cm,width=7.5cm]{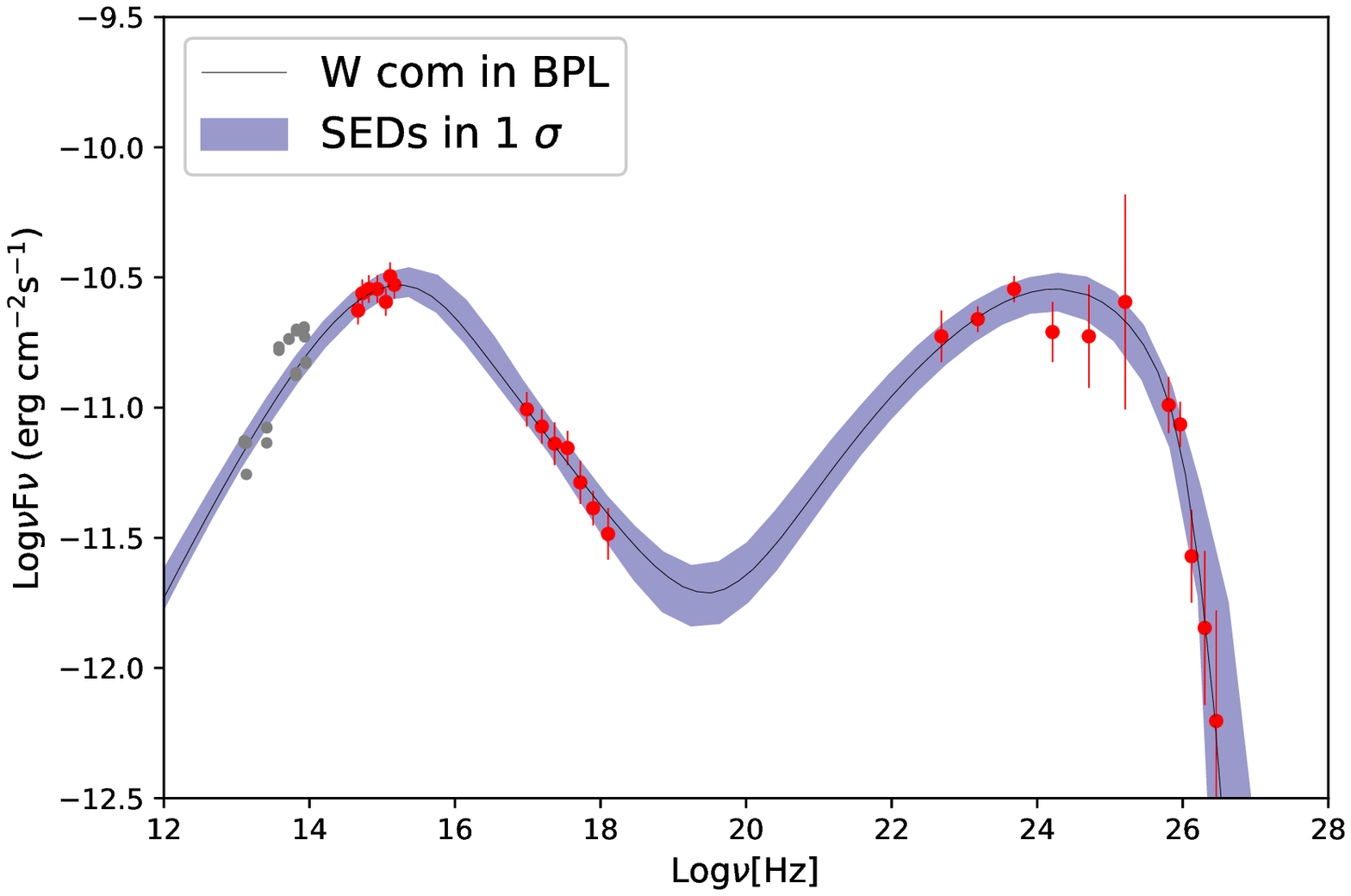}
%\caption{fig1}
}
\quad
\subfigure[]{
\includegraphics[height=6.cm,width=7.5cm]{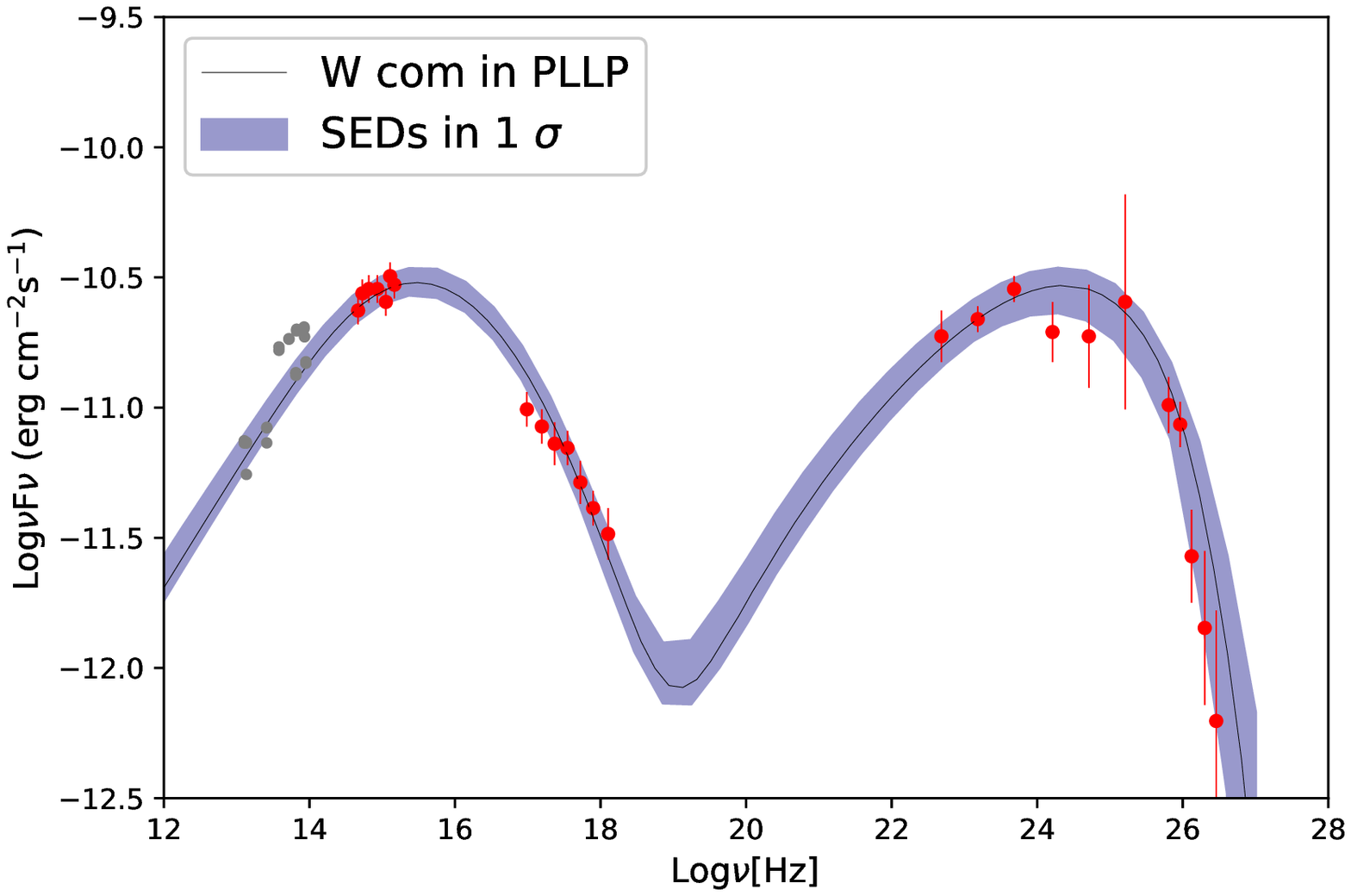}
}
\flushright
%\centering
\quad\quad\quad\quad\quad\quad
\subfigure[]{
\includegraphics[height=5.8cm,width=7.cm]{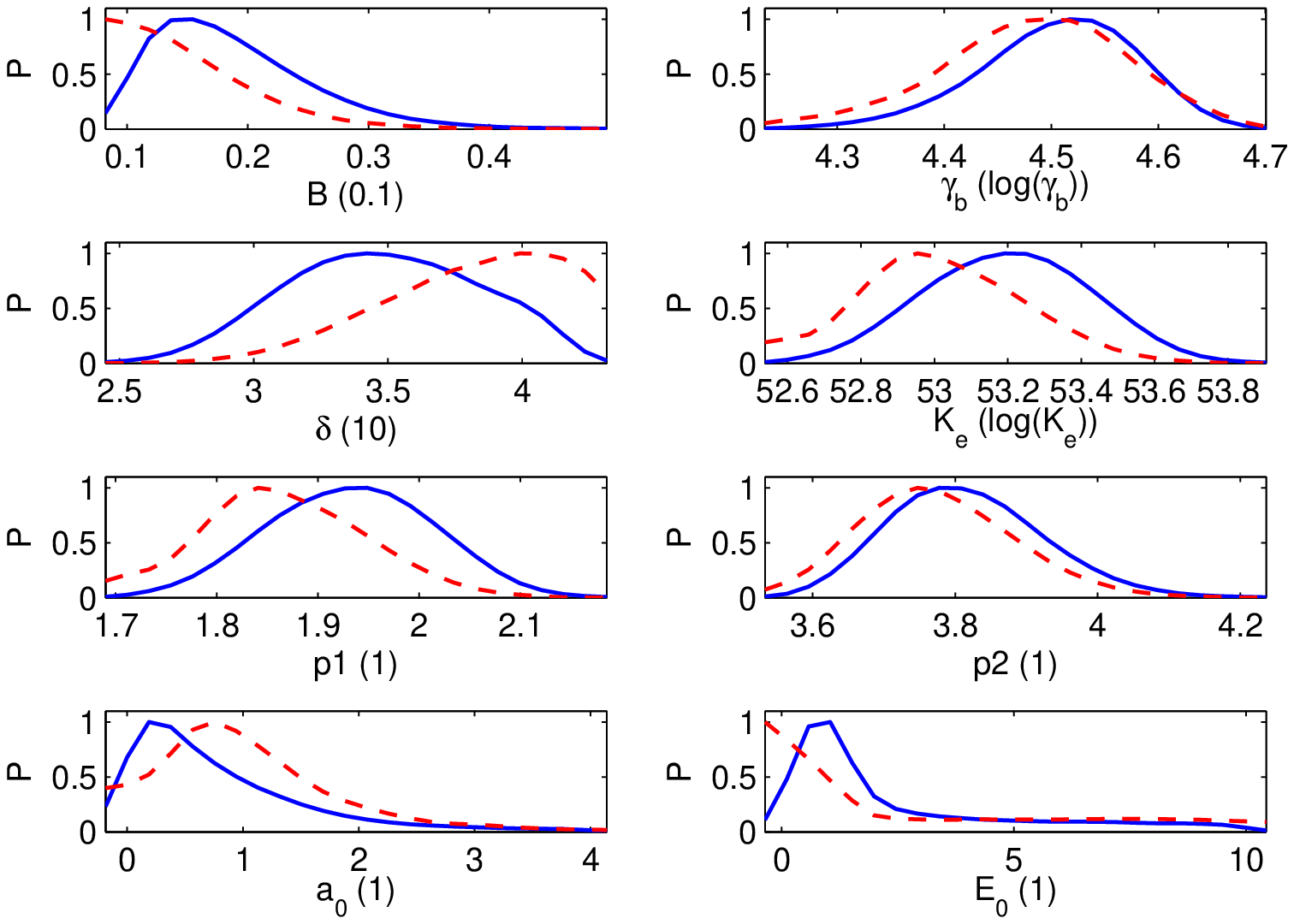}
}
\quad\quad\quad\quad
\subfigure[]{
\includegraphics[height=5.8cm,width=7.cm]{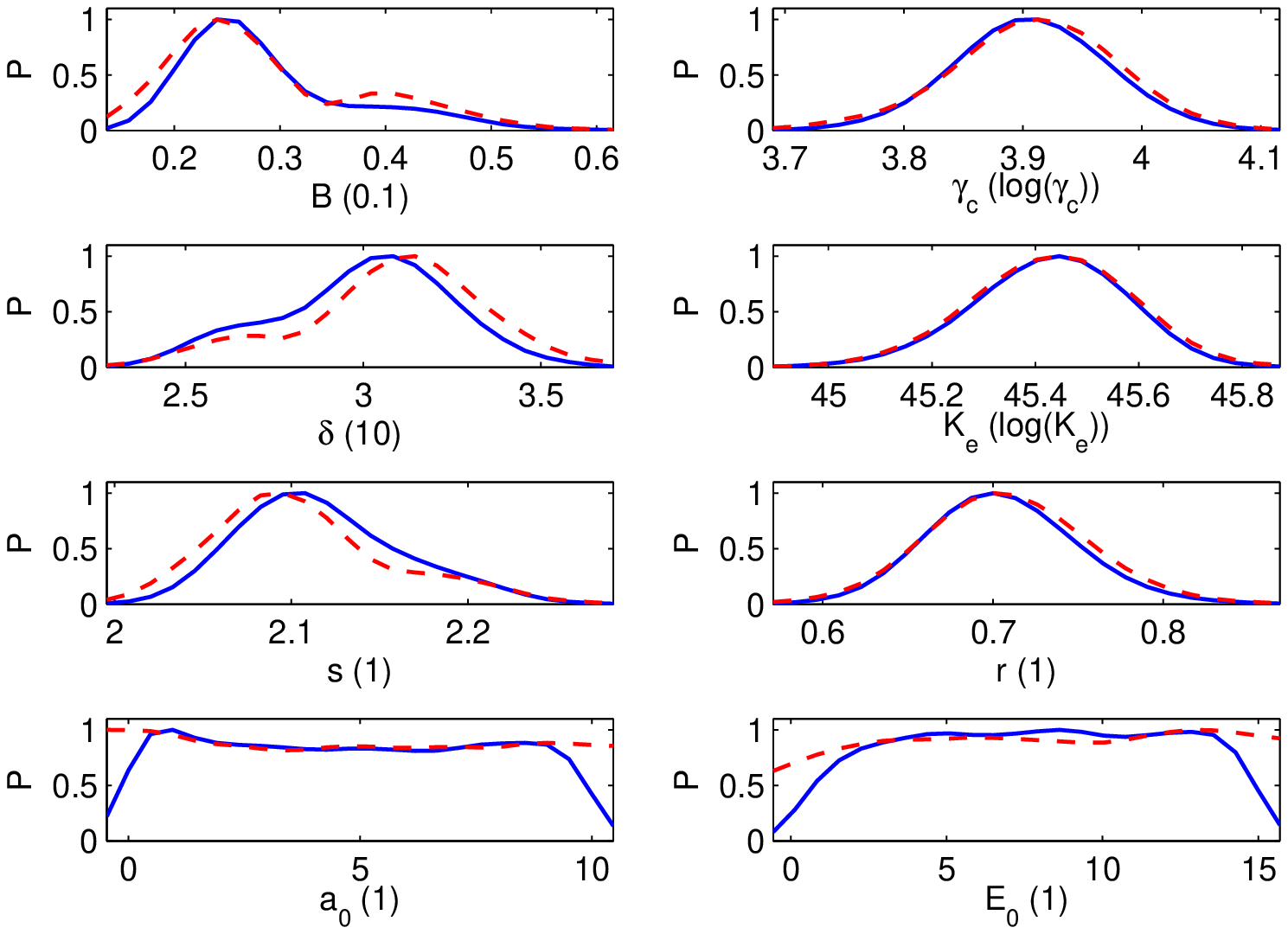}
}
\caption{As Fig. 1, but for W com.}
\end{figure*}
     		
\begin{figure*}
%\centering
\flushright
\subfigure[]{
\includegraphics[height=6.cm,width=7.5cm]{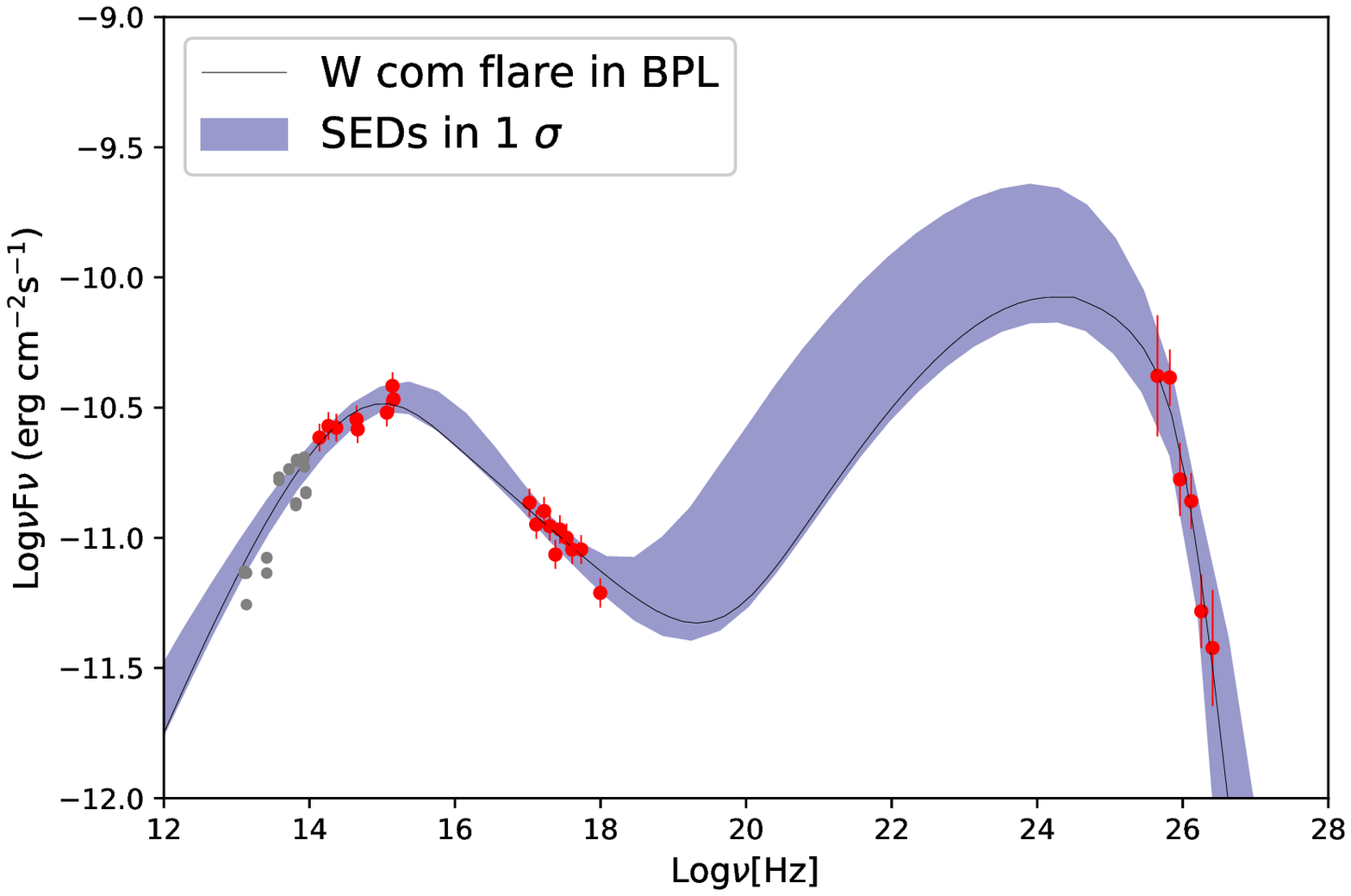}
%\caption{fig1}
}
\quad
\subfigure[]{
\includegraphics[height=6.cm,width=7.5cm]{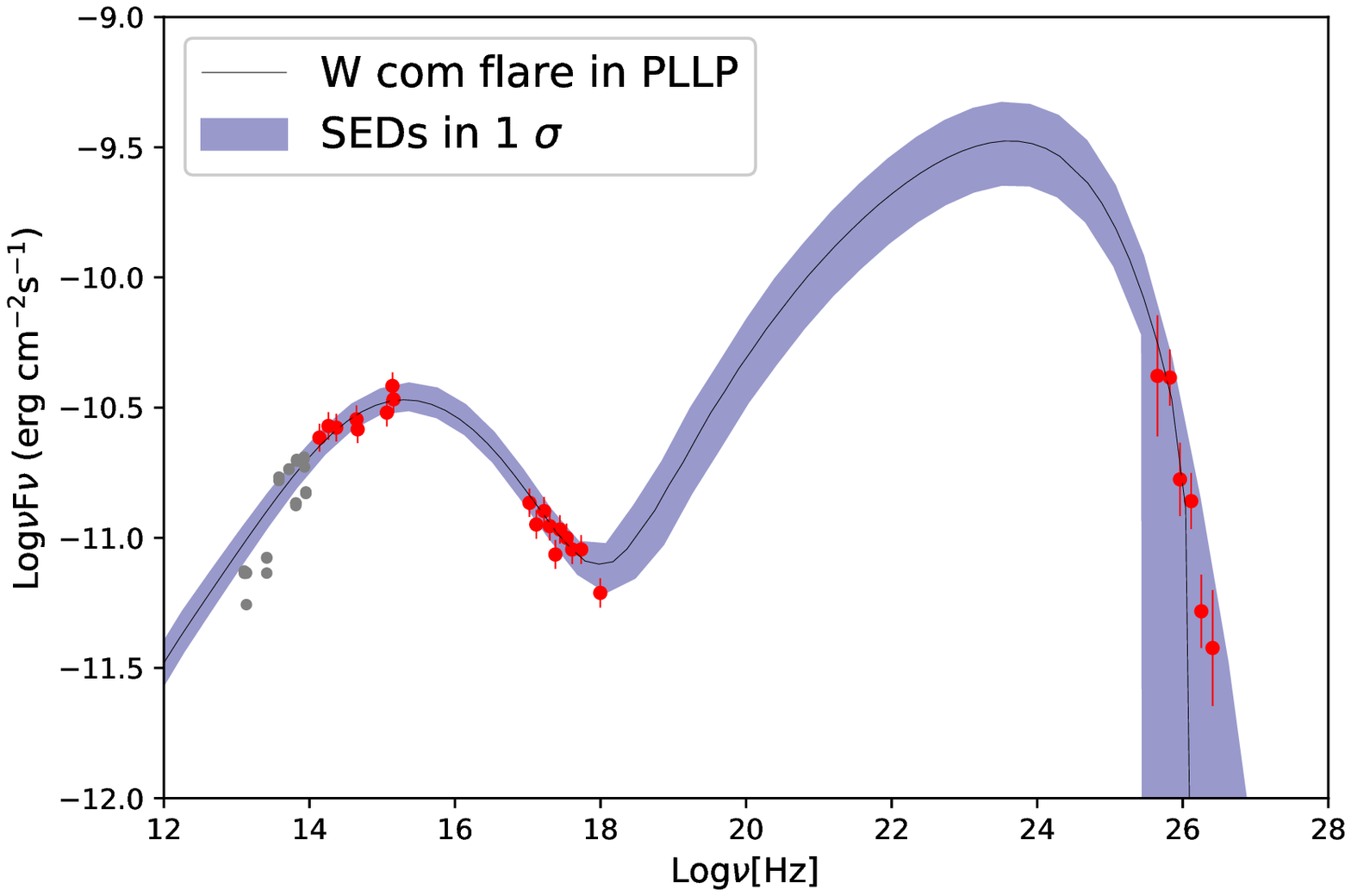}
}
\flushright
%\centering
\quad\quad\quad\quad\quad\quad
\subfigure[]{
\includegraphics[height=5.8cm,width=7.cm]{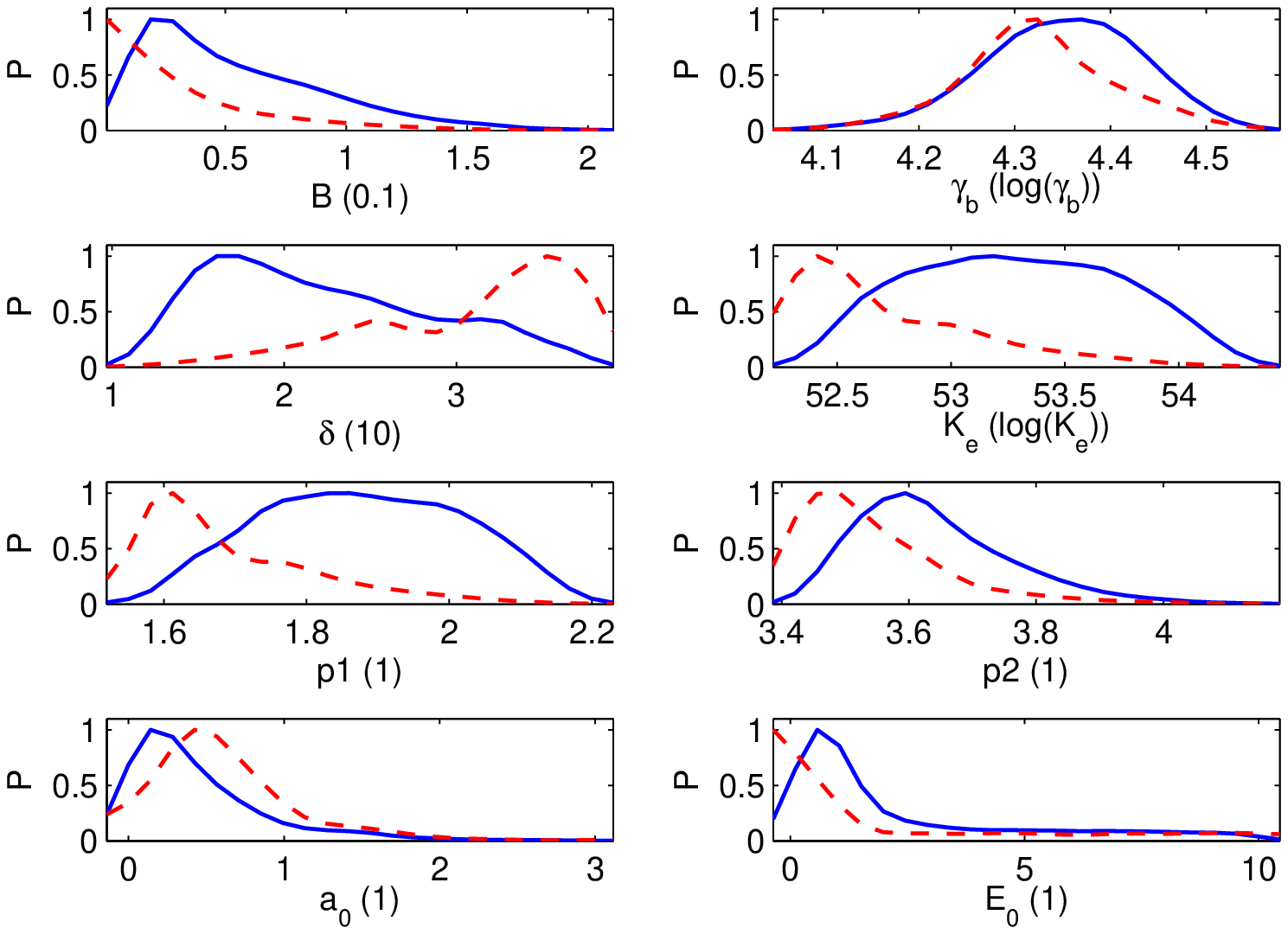}
}
\quad\quad\quad\quad
\subfigure[]{
\includegraphics[height=5.8cm,width=7.cm]{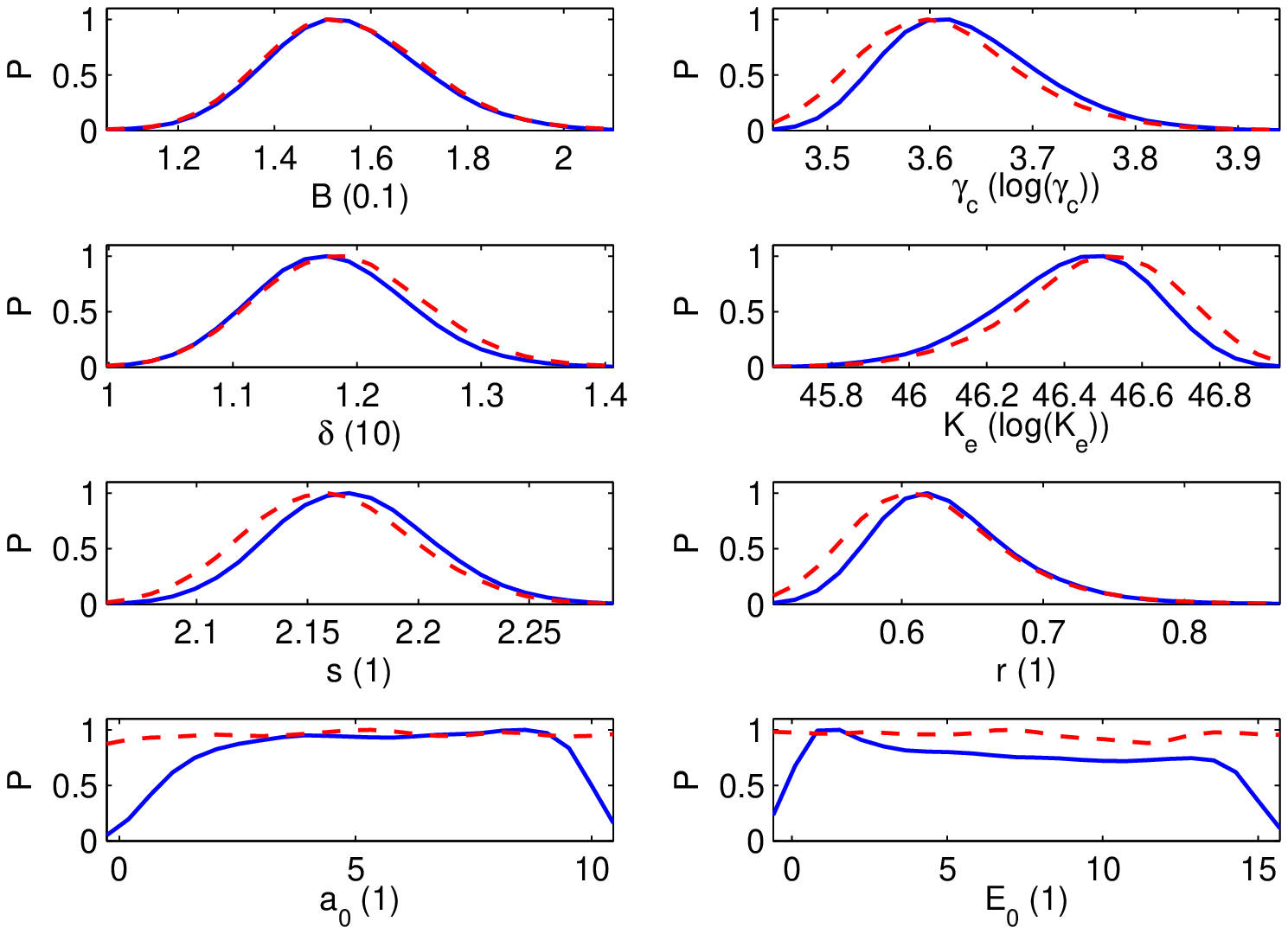}
}
\caption{As Fig. 1, but for W com in the flare stage.}
\end{figure*}

%end the fitting figure
%EBL::::::::::::::::::
\begin{figure*}
\centering
\subfigure{\includegraphics[height=6.5cm,width=8.7cm]{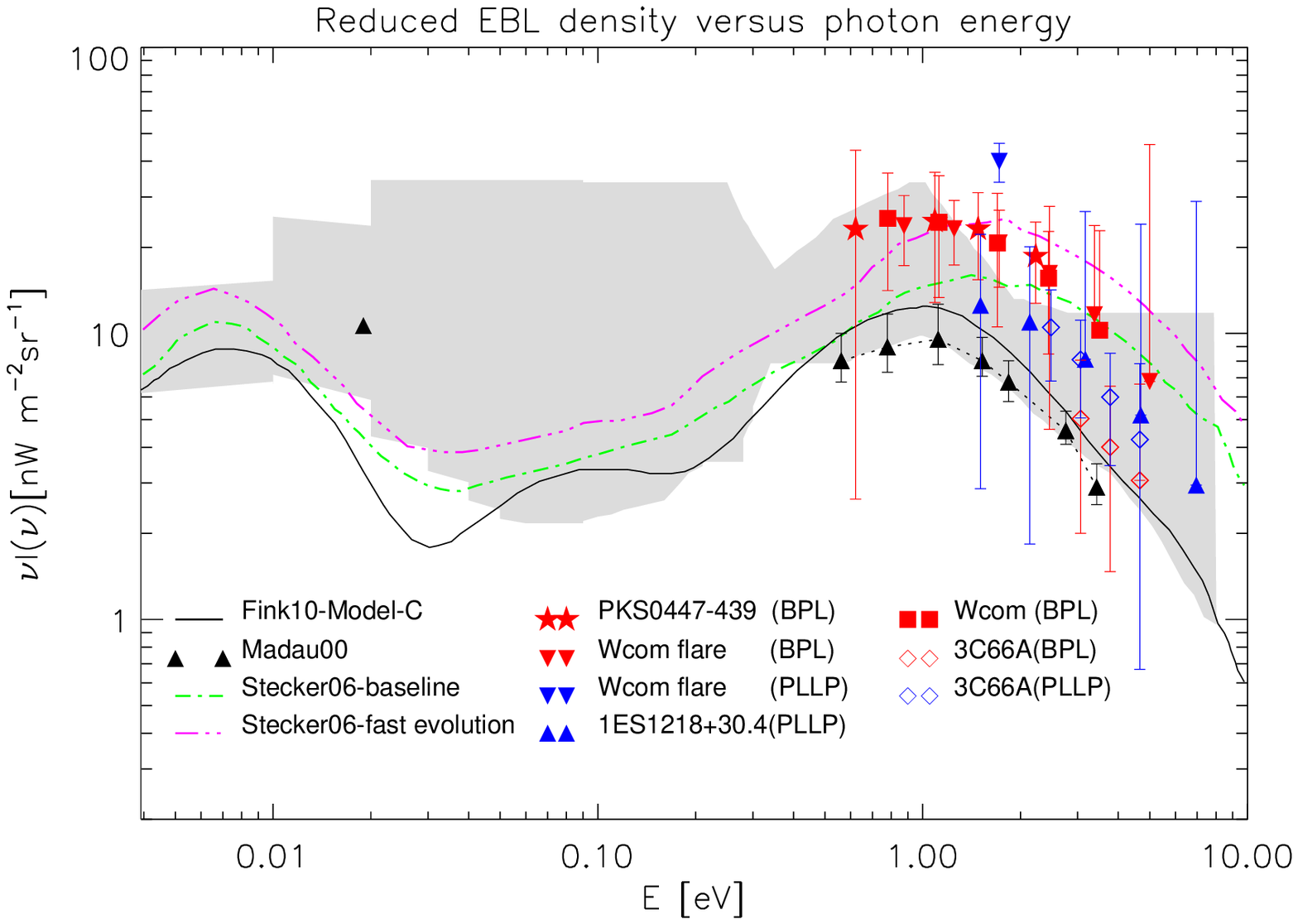}}
\subfigure{\includegraphics[height=6.5cm,width=8.7cm]{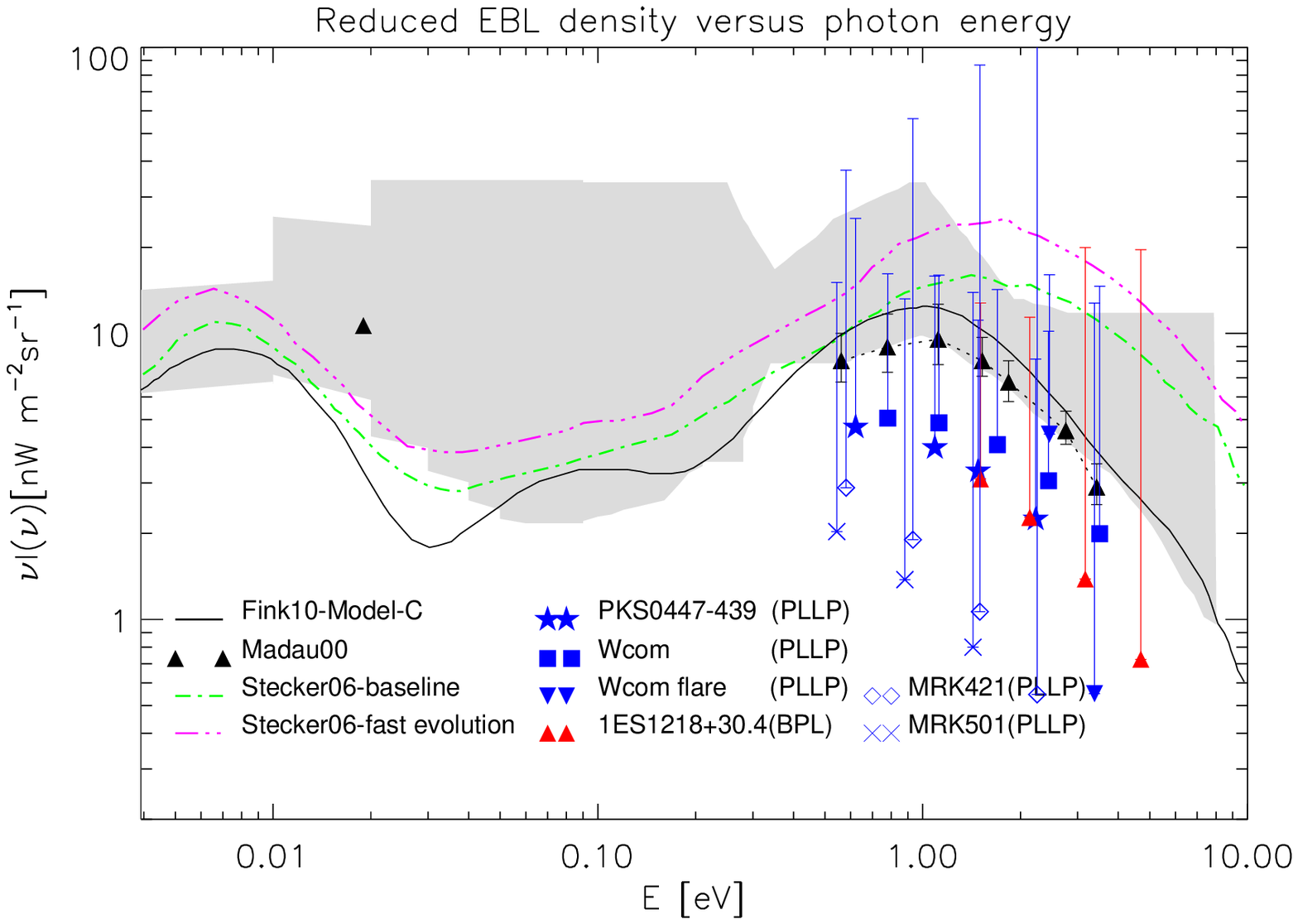}}
\caption{Constrains of EBL by TeV BL Lacs. several EBL models are plotted: the
baseline and fast evolution models of \emph{Stecker06} \citep{ste2006}, the \emph{Finke 10-Model-C} \citep{fin2010} and the lower limits of EBL of \emph{Madau00} from \citet{mad2000}. The shaded area
represents the range of the allowed EBL intensity as determined by UV to sub millimeter observations. The left and right panel depict that the EBL densities are larger and lower than the value of \emph{Madau00}, respectively}
\end{figure*}

\section{DISCUSSION AND CONCLUSION}

In this paper, we present the study of the EBL density by fitting observed SEDs of TeV BL Lacs objects based on the semi-model dependent optical depth. However, we should point out that, in our method, three issues should be taken into account. 1) The reliable intrinsic spectra. We use one-zone SSC to fit the SEDs, which has already been widely used in several paper. Meanwhile, we take two types of EEDs to govern the SEDs for finding the change of the EBL spectrum for the uncertain EEDs in the emitting region. Then we also use (quasi-) simultaneous multiwavelength data to guarantee that the SSC mechanism is appropriated. Besides, several radiation mechanisms like the hadronic model still have problem, which require extremely physical condition such as large magnetic field and plentiful protons. 2) The EBL optical depth temple. As discussed in the introduction, we offer a more flexible optical depth model to fit the SEDs with a few parameters and can get better result. 3) The fitting method. Unlike "eyeball fitting", we use the MCMC method to explore the parameter space during our fitting processes.  

As comparisons, we also plot the results of some EBL models in the two panels of Fig.8, including the  models of \cite{ste2006},  \cite{fin2010}, and the lower limits of EBL of \emph{Madau00} from \cite{mad2000}. In the left panel, we plot the EBL density that is higher than the lower EBL one offered by \cite{mad2000}. In the right panel, we find that the EBL density is lower than the limit given by galaxy counts \citep{mad2000}.

In general, the results we generated are plausible, in which the EBL densty is roughly consisted with that determined by UV to sub millimeter observations. We have shown that the upper limit of EBL densities are close to 1~-~30 n W m$^{-2}$ sr$^{-1}$ for the EBL photons with 1 eV$\sim$ 5 eV, which are similar to the published measurement. In addition, it is found that our result are more likely consistent with the baseline and fast evolution models of \citep{ste2006}, which the EBL density is about 20 n W m$^{-2}$ sr$^{-1}$. Note that the EBL density obtained by lower redshift sources are much closer to the previous results than that reduced by high redshift objects, which disfavor that the EBL is more transparent to distant $\gamma$-ray \citep {fra2019}; Besides, for lower redshift objects, Mrk 421 and Mrk 501, they seemly fail to obtain reasonable limits on the EBL density. As shown in Fig.8, both in BPL and PLLP EEDs, the several reduced EBL densities are lower than those obtained by galaxy counts. It is implied that the one-zone SSC model could not explain those TeV objects, and other radiation components from Hadronic process \citep{zheng2016} or electromagnetic cascades (EM) \citep{chen2015} should be considered. \cite{yan2013} have used the different EEDs in the frame of a one-zone SSC model to investigate the acceleration processes in the jet of BL Lacs by analyzing the fitting SEDs. However, it is still hard to find out the discrepancy among those spectra for the limit of observation. Interestingly, our method can be used to test the radiation mechanism by simply comparing the reduced EBL density with the lower limits from galaxy counts. Furthermore, the lower EBL density in this work may be reasonable, because non-standard effects, such as the axion-like particles (ALPs) \cite{arl2014} and violation of Lorentz invariance (LIV) \cite{aka2010}, may allow the VHE spectra to void the EBL absorption.

\begin{table}
%\centering
\small
%\begin{center}
\caption{The equipartition parameter between the comoving
electron and magnetic field energy density }
\label{tab-3}
\setlength{\tabcolsep}{7mm}{
\begin{tabular}{lcl}
\hline
\hline
Source name  &EEDs&   $u'_e/u'_B$  \\
~[1]          &[2] &   ~~[3]   \\
\hline
\hline
 1ES1218+30.4&BPL &   10.3  \\
 ~           &PLLP&   486.4 \\    
 3C66A       &BPL &   33.3 \\  
 ~           &PLLP&   47.1 \\  
 MRK421      &BPL &   35.3 \\  
 ~           &PLLP&   33.2  \\  
 MRK501      &BPL &   22.7  \\  
 ~           &PLLP&   16.9  \\   
 PKS0447-439 &BPL &   41.7  \\   
 ~           &PLLP&   33.5  \\   
 Wcom        &BPL &   595.5 \\   
 ~           &PLLP&   180.4 \\    
 Wcom flare  &BPL &   1607.8\\     
 ~           &PLLP&   208.8 \\ 

\hline
\end{tabular}}
%\end{center}
\end{table}
   
  Besides, after our fitting process, we also calculate the ratio between the comoving electron and magnetic field energy density as $u'_e/u'_B=\frac{m_ec^2\int{\gamma}'n'(\gamma')d{\gamma}'}{B^2/8\pi}$, where $n'(\gamma')=\frac{3N(\gamma')}{4\pi {R'_b}^3}$ is electron number density per volume. As shown in Table 3, our result indicts that most sources are near equipartition, and the value of $u'_e/u'_B$ is cluster at 40, favoring the HBL scenario that those BL Lacs need not to introduce a additional VHE component \citep{der2014}. Comparing our discussion above, it is not contradictory, for one reason,  our fitting is suitable, or for another, the VHE component may exist and be weaked by our flexible optical depth model. In addition, from the Talbe 1 and 2, the Doppler factors $\delta_{\rm {D}}$ and the magnetic field $B$ are clustered at 25, and 0.01, respectively, showing that the TeV objects in our paper are "normal".

As shown in Fig.8, the $\gamma$-ray observation errors are still large, leading to the huge uncertain in the reduced EBL density. In addition, the EEDs, commonly used to model the SEDs \cite{yan2013,zhou2014,qin2018a}, need further to be tested. The recent investigation \citep{zheng2018} suggested that the EED has many different shapes if the stochastic Fermi acceleration, radiative losses, particle injection and escape are considered. Besides, as discussed above, before using TeV observation to study the EBL density, we should make sure that the (quasi-) simultaneous multi-waveband SEDs are needed and one-zone SSC model is appropriated for certain TeV sources.  Furthermore, the monochromatic assumption in this work could lead to a higher upper limit, as discussed by \citep{fin2009}.  
%______________________________________________________________
\section*{Acknowledgments}

We thank the anonymous referee and the editor for useful comments and suggestions. This research has made use of the NASA/IPAC Extragalactic Database (NED) which is operated by the Jet Propulsion Laboratory, California Institute of Technology, under contract with the National Aeronautics and Space Administration. The authors gratefully acknowledge the financial supports from the National Natural Science Foundation of China (grants 11663008, 11661161010 and 11802265), the Science Research Foundation of Yunnan Education Department of China (grants 2017ZZX177, 2018JS422, 2019J0733, and 2019J0734) and Yunnan local colleges applied basic research projects (grants 2019FH001-012, 2019FH001-076). The author (QLH) gratefully acknowledge the financial supports from the Hundred Talents Program of Yuxi (grants 2019). The authors gratefully acknowledge the computing time granted by the Yuxi normal university and provided on the facilities at the Yuxi normal university Supercomputing Platform.

\section*{Data availability}
The data underlying this article is from references compiled by \cite{zhang2012}, and will be shared on reasonable request to the corresponding author.

%%\clearpage

%++++++++++++++++++++++++++++++++++++++++++++++++++++++++++++++++++++++++++++++++

%\end{document}
\clearpage

\begin{thebibliography}{}
\bibitem[Acciari et al.(2010)]{acc2010} Acciari, V.~A., et al.\
2010, ApJL, 708, L100
\bibitem[Acciari et al.(2019)]{acc2019} Acciari, V.~A., Ansoldi, S., Antonelli, L.~A., et al.\ 2019, MNRAS, 486, 4233
\bibitem[Ackermann et al.(2012)]{ack2012} Ackermann, M., Ajello, M., Allafort, A., et al.\ 2012, Science, 338, 1190
\bibitem[Aharonian et al.(1999)]{aha1999} Aharonian, F.~A., Akhperjanian, A.~G., Barrio, J.~A., et al.\ 1999, A\&A, 349, 11
\bibitem[{{Aharonian} {et~al.}(2006a)}]{aha2006}
---. 2006a, Nature, 440, 1018
\bibitem[Akahori, \& Ryu(2010)]{aka2010} Akahori, T., \& Ryu, D.\ 2010, ApJ, 723, 476
\bibitem[Arlen et al.(2014)]{arl2014} Arlen, T.~C., Vassilev, V.~V., Weisgarber, T., et al.\ 2014, ApJ, 796, 18
\bibitem[Band et al.(1993)]{band1993} Band, D., Matteson, J., Ford, L., et al.\ 1993, ApJ, 413, 281
\bibitem[Becker et al.(2006)]{bec2006} Becker, P.~A., Le, T., \& Dermer, C.~D.\ 2006, ApJ, 647, 539
\bibitem[Bernstein et al.(2002)]{ber2002} Bernstein, R.~A., Freedman, W.~L., \& Madore, B.~F.\ 2002, ApJ, 571, 107
\bibitem[Chen et al.(2015)]{chen2015} Chen, W., Buckley, J.~H., \& Ferrer, F.\ 2015, Phys. Rev. Lett., 115, 211103
\bibitem[Desai et al.(2019)]{des2019} Desai, A., Helgason, K., Ajello, M., et al.\ 2019, ApJL, 874, L7
\bibitem[Dermer et al.(2009)]{der2009} Dermer, C.~D., Finke, J.~D., Krug, H., et al.\ 2009, ApJ, 692, 32
\bibitem[Dermer et al.(2014)]{der2014} Dermer, C.~D., Cerruti, M., Lott, B., et al.\ 2014, ApJ, 782, 82
\bibitem[Dom{\'\i}nguez et al.(2013)]{dom2013} Dom{\'\i}nguez, A., Finke, J.~D., Prada, F., et al.\ 2013, ApJ, 770, 77
\bibitem[Dwek \& Slavin(1994)]{dewk1994} Dwek, E., \& Slavin, J.\ 1994, ApJ, 436, 696
\bibitem[Driver et al.(2016)]{dri2016} Driver, S.~P., Andrews, S.~K., Davies, L.~J., et al.\ 2016, ApJ, 827, 108
\bibitem[Fazio et al.(2004)]{faz2004} Fazio, G.~G., Ashby, M.~L.~N., Barmby, P., et al.\ 2004, ApJS, 154, 39
\bibitem[Finke et al.(2008)]{fin2008} Finke, J.~D., Dermer, C.~D., B{\"o}ttcher, M.\ 2008, ApJ, 686, 181-194
\bibitem[Finke \& Razzaque(2009)]{fin2009} Finke, J.~D., \& Razzaque, S.\ 2009, ApJ, 698, 1761
\bibitem[Finke et al.(2010)]{fin2010} Finke, J.~D., Razzaque, S., \& Dermer, C.~D.\ 2010, ApJ, 712, 238
\bibitem[{{Franceschini} {et~al.}(2008)}]{fra2008} {Franceschini}, A., {Rodighiero}, G., \& {Vaccari}, M. 2008, A\&A, 487, 837
\bibitem[Franceschini et al.(2019)]{fra2019} Franceschini, A., Foffano, L., Prandini, E., et al.\ 2019, A\&A, 629, A2
\bibitem[Fossati et al. (2008)]{fos2008} Fossati, G. et al. 2008, ApJ, 677, 906
\bibitem[Ghisellini et al.(1998)]{Ghi1998} Ghisellini, G., Celotti, A., Fossati, G., Maraschi, L., \& Comastri, A.\ 1998, MNRAS, 301, 451
\bibitem[Ghisellini \& Tavecchio(2010b)]{Ghi2010b} Ghisellini, G., \& Tavecchio, F.\ 2010, MNRAS, 409, L79
\bibitem[{{Gilmore} {et~al.}(2009)}]{gil2009}Gilmore, R.C., Madau, P., Primack, J.R.,
Somerville, R.S., \& Haardt, F., 2009, MNRAS, 399, 1694
\bibitem[Hauser(1998)]{hau1998} Hauser, M.~G.\ 1998, American Astronomical Society Meeting Abstracts 193, 62.02.
\bibitem[Inoue \& Tanaka(2016)]{ino2016} Inoue, Y., \& Tanaka, Y.~T.\ 2016, ApJ, 828, 13
\bibitem[{{Gould} \& {Schr{\'e}der}(1967)}]{gou1967}{Gould}, R.~J. \& {Schr{\'e}der}, G.~P. 1967, Physical Review, 155,1404
\bibitem[{{Kneiske} {et~al.}(2004)}]{kne2004} {Kneiske}, T.~M., {Bretz}, T., {Mannheim}, K., \& {Hartmann}, D.~H. 2004, A\&A,
  413, 807
\bibitem[{{Kneiske} {et~al.}(2002){kne2002}, {Mannheim}, \&  {Hartmann}}]{kneiske02}
{Kneiske}, T.~M., {Mannheim}, K., \& {Hartmann}, D.~H. 2002, A\&A, 386, 1
\bibitem[Lewis \& Bridle(2002)]{lew2002} Lewis, A., \& Bridle, S.\ 2002, Phys. Rev., 66, 103511
\bibitem[Mackay(2003)]{mac2003} Mackay, D.~J.~C.\ 2003, Information Theory, Inference and Learning Algorithms, UK: Cambridge University Press, 640
\bibitem[{{Madau} \& {Pozzetti}(2000)}]{mad2000} {Madau}, P. \& {Pozzetti}, L. 2000, MNRAS, 312, 9
\bibitem[{{Mazin} \& {Raue}(2007)}]{maz2007} {Mazin}, D. \& {Raue}, M
\bibitem[Moderski et al.(2005)]{mod2005} Moderski, R., Sikora, M., Coppi, P.~S., et al.\ 2005, MNRAS, 363, 954
\bibitem[Prandini et al.(2012)]{pra2012} Prandini, E., Bonnoli, G., \& Tavecchio, F.\ 2012, A\&A, 543, A111
\bibitem[Qin et al.(2018a)]{qin2018a} Qin, L., Wang, J., Yan, D., et al.\ 2018, MNRAS, 473, 3755
\bibitem[Qin et al.(2018b)]{qin2018b} Qin, L., Wang, J., Yang, C., et al.\ 2018, PASJ, 70, 5
\bibitem[{{Schroedter}(2005)}]{sch2005} {Schroedter}, M. 2005, ApJ, 628, 617
\bibitem[Sinha et al.(2014)]{sin2014} Sinha, A., Sahayanathan, S., Misra, R., et al.\ 2014, ApJ, 795, 91
\bibitem[{{Stanev} \& {Franceschini}(1998)}]{sta1998} {Stanev}, T. \& {Franceschini}, A. 1998, ApJ, 494, L159
\bibitem[{{Stecker} \& {de Jager}(1993)}]{ste1993} {Stecker}, F.~W. \& {de Jager}, O.~C. 1993, ApJ, 415, L71
\bibitem[{{Stecker} {et~al.}(2006)}]{ste2006} {Stecker}, F.~W., {Malkan}, M.~A., \& {Scully}, S.~T. 2006, ApJ,
648, 774
\bibitem[Tavecchio et al.(2010)]{tav2010} Tavecchio, F., Ghisellini, G., Ghirlanda, G., et al.\ 2010, MNRAS, 401, 1570
\bibitem[Tramacere et al.(2011)]{tra2011} Tramacere, A., Massaro, E., \& Taylor, A.~M.\ 2011, ApJ, 739, 66
\bibitem[Urry \& Padovani (1999)]{urr1999} Urry, C. M. \& Padovani, P. 1999, ApJ, 11, 159
\bibitem[Xie et al.(2001)]{xie2001} Xie, G.~Z., Li, K.~H., Bai, J.~M., et al.\ 2001, ApJ, 548, 200
\bibitem[Yan et al.(2013)]{yan2013} Yan, D., Zhang, L., Yuan, Q., Fan, Z., \& Zeng, H.\ 2013, ApJ, 765, 122
\bibitem[Yang, \& Wang(2010)]{yang2010} Yang, J., \& Wang, J.\ 2010, A\&A, 522, A12
\bibitem[Zhang et al. (2012)]{zhang2012} Zhang, J., Liang, E. W., Zhang, S. N., \& Bai, J. M. 2012, ApJ, 752, 157
\bibitem[Zheng et al.(2016)]{zheng2016} Zheng, Y.~G., Yang, C.~Y., \& Kang, S.~J.\ 2016, A\&A, 585, A8
\bibitem[Zheng et al.(2018)]{zheng2018} Zheng, Y.~G., Long, G.~B., Yang, C.~Y., et al.\ 2018, PASJ, 130, 083001
\bibitem[Zhou et al.(2014)]{zhou2014} Zhou, Y., Yan, D., Dai, B., \& Zhang, L.\ 2014, PASJ, 66, 12
\bibitem[Zhu et al.(2016)]{zhu2016} Zhu, Q., Yan, D., Zhang, P., et al.\ 2016, MNRAS, 463, 4481
\end{thebibliography}
\end{document}